\documentclass[onecollarge,final]{svjour2}

\smartqed

\usepackage [dvips]{graphicx}
\usepackage{amssymb}
\usepackage{epsfig}
\usepackage{graphicx}
\usepackage{verbatim}

\usepackage[T1]{fontenc}

\usepackage{amsmath}
\usepackage{enumerate}
\usepackage{graphicx}
\usepackage{mathbbol}
\usepackage{amsfonts}

\usepackage{graphicx,hyperref}
\usepackage{mathtools}
\usepackage{amsfonts}
\usepackage{bbold}
\usepackage{bm}
\usepackage{multirow}
\usepackage{xcolor}
\usepackage{tikz}
\usetikzlibrary{shapes,shapes.multipart}

\begin{document}


\title{Maximal distance travelled by $N$ vicious walkers till their survival}



\author{Anupam Kundu \and Satya N. Majumdar \and Gr\'egory Schehr}

\institute{A. Kundu \at  Laboratoire de Physique Th\'eorique et Mod\`eles
  Statistiques, Universit\'e Paris-Sud, B\^at. 100, 91405 Orsay Cedex, France. \email{anupamoraja@gmail.com} \\  \and S.~N. Majumdar \at Laboratoire de Physique Th\'eorique et Mod\`eles
  Statistiques, Universit\'e Paris-Sud, B\^at. 100, 91405 Orsay Cedex,
France. \email{satya.majumdar@u-psud.fr} \\ \and G. Schehr  \at Laboratoire de Physique Th\'eorique et Mod\`eles
  Statistiques, Universit\'e Paris-Sud, B\^at. 100, 91405 Orsay Cedex, France. \email{gregory.schehr@u-psud.fr}}

\date{\today}

\maketitle

\begin{abstract}
We consider $N$ Brownian particles moving on a line starting from initial positions 
${\bf{u}}\equiv \{u_1,u_2,\dots u_N\}$ such that $0<u_1 < u_2 < \cdots < u_N$. Their motion gets stopped at time $t_s$ when either 
two of them collide or when the particle 
closest to the origin hits the origin for the first time. For $N=2$, we study the probability distribution function $p_1(m|{\bf{u}})$ and $p_2(m|{\bf{u}})$ of the 
maximal distance travelled by the $1^{\text{st}}$ and $2^{\text{nd}}$ walker till $t_s$. For general $N$ particles with identical diffusion constants $D$, we show that
the probability distribution $p_N(m|{\bf u})$ of the global maximum $m_N$, has a power law tail 
$p_i(m|{\bf{u}}) \sim {N^2B_N\mathcal{F}_{N}({\bf u})}/{m^{\nu_N}}$ with exponent $\nu_N =N^2+1$. We obtain explicit expressions of the function  
$\mathcal{F}_{N}({\bf u})$ and of the $N$ dependent amplitude $B_N$ which we also analyze for large $N$ using techniques from random matrix theory. We verify our analytical results 
through direct numerical simulations.
\end{abstract}
\maketitle

\section{Introduction}

Extreme value statistics (EVS) is by now a major issue with a variety of applications in several areas of sciences including 
physics, statistics or finance, to name just a few~\cite{Gumbel}. For $N$ independent 
and identically distributed (i.i.d.) random variables $y_1, \cdots, y_N$, the distribution of the maximum 
$y_{\rm max} = \max_{1 \leq i \leq N}\, y_i$ (or the minimum $y_{\rm min}$) 
is well understood with the identification, in the large $N$ (thermodynamical) limit, of three distinct universality classes, 
depending on the parent distribution of the $y_i$'s \cite{Gumbel}. However, these results for i.i.d. random variables do not apply 
when the random variables are correlated  \cite{Dean,KrapivskySatya2003}. Recently, there has been a surge of interest in EVS of 
{\it strongly correlated} random variables, which is very often the interesting case in statistical physics. Physically relevant 
examples include for instance the extreme statistics of a stochastic process $u(t)$, with strong temporal correlations, 
like Brownian motion or its variants. Many studies in this context are focused on extremal properties, like the maximum of $u(t)$, 
over a {\it fixed} time interval, $t \in [0,T]$~\cite{Revuz99,Yor01,Borodin02,Majumdar04,Majumdar05,satyacurrsci,Schehr10}.  

However in many cases the length of this time interval is itself a random variable $t_s$, which can thus vary from one realization of the 
stochastic process to another. This time $t_s$ is usually strongly correlated to the process $u(t)$ itself. 
An interesting situation is the case where 
$t_s$ is a ``stopping time'' \cite{Feller}, i.e. when it is associated to the stopping of the process $u(t)$ if a certain event occurs for the 
first time.  For example, in a queuing process starting from an initial 
queue length $l_0 > 0$,   
$t_s$ is the time 
 when the queue length $l_t$ becomes zero for the first time (also called the ``busy period'') \cite{Kearney04,Kearney06}. 
In finance $t_s$ might correspond to 
the time when a stock price $S_t$ reaches some specified level for the first time \cite{Yor01,Comtet98,majum-Qfin}. Stopping times also 
naturally arise in various statistical physics
models ranging from capture processes \cite{Bramson91,Wenbo01} or target annihilating problems \cite{Blumen84} all the way  
to reaction-diffusion kinetics \cite{Kang85,Ben-Naim93,Krapivsky95} or coarsening dynamics of domain walls in Ising model \cite{Bray94}. 
In the context of stochastic control theory, stochastic processes with ``stopping time'' have been widely studied~\cite{Shiryaev}.

The simplest example of a ``stopped'' stochastic process is the motion of a single Brownian particle starting form $u(0)=u_1>0$
which is observed till time $t_s$ when the walker crosses the origin for the first time. 
This time is called the first passage time \cite{Rednerbook,Bray13}. 
A natural extreme value question is then: what is the distribution $p(m|u_1)$ of the maximal displacement $m=\max_{0\leq t\leq t_s}u(t)$ 
travelled by the walker till its first passage time $t_s$ ? It can be shown \cite{Kearney05} that the cumulative probability 
$Q_1(u_1|L)=\text{Prob}[m\leq L |u_1]$ that the maximum stays below $L$ till the first passage time is given by 
$Q_1(u_1|L)=1-{u_1}/{L}$, hence $p(m|u_1)={u_1}/{m^2}$. In the context of polymer translocation through a small pore, the quantity $1-Q_1(u_1|L)$ is 
precisely the probability of complete translocation of a polymer of length $L$. For generic subdiffusive process, 
this translocation probability is shown to scale as $\sim (u_1/L)^{\phi}$ for large $L$ with $\phi=\theta_p/H$ where $\theta_p$ is the persistence 
exponent \cite{Bray13,Satyacurrsci99} and $H$ is the Hurst exponent \cite{Satyarosso10}. 
Other related questions like the statistics of the time when the walker reaches the maximal displacement before its first 
passage time $t_s$ or the fluctuations of the area enclosed under the Brownian motion till $t_s$,  
have also been studied in connection with several applications including queuing theory or lattice polygon models 
\cite{Kearney04,Kearney06,Kearney05,Kearney07,Randon-Furling07,Abundo13}.

``Stopped'' processes involving $N > 1$ particles are also interesting and have been considered in the literature. For instance, the maximal 
displacement between the ``leader" and the ``laggard" among $N$ particles has been studied for $N=3$ particles in Ref. \cite{SatyaBray2010}. 
Very recently the authors of Ref. \cite{KrapivskySatya2010} have studied the probability distribution function (PDF) $p(m|{\bf{u}})$ of the 
global maximum $m_N$ of $N$ non-interacting and identical Brownian walkers (i.e. with the same diffusion constant) before their first exit from 
the positive half-line, given that they had started from positions ${\bf{u}}\equiv \{u_1,u_2,\dots u_N\}$. 
They showed that the tail of the PDF $p(m|{\bf{u}})$ 
of the global maximum $m_N$ till the time $t_s$ when any one of the $N$ walkers crosses the origin for the first time, is given by 
\begin{equation}
p(m|{\bf{u}}) \simeq \left [N~b_N \prod_{i=1}^N u_i\right ]~\frac{1}{m^{N+1}} \;,\; m\gg u_N \label{non_int}   \;,
\end{equation}
where $b_N$ is an $N$-dependent constant that behaves for large $N$ as, $b_N \approx \text{exp}\left[\frac{N}{2}\log (\log N)\right]$. 
This result (\ref{non_int}) holds for non-interacting particles and it is natural to wonder
about the effects of interactions on the statistics of the global maximum till the stopping time of this multi-particle process. 

\begin{figure}[t]
\centering
\includegraphics[scale=0.5]{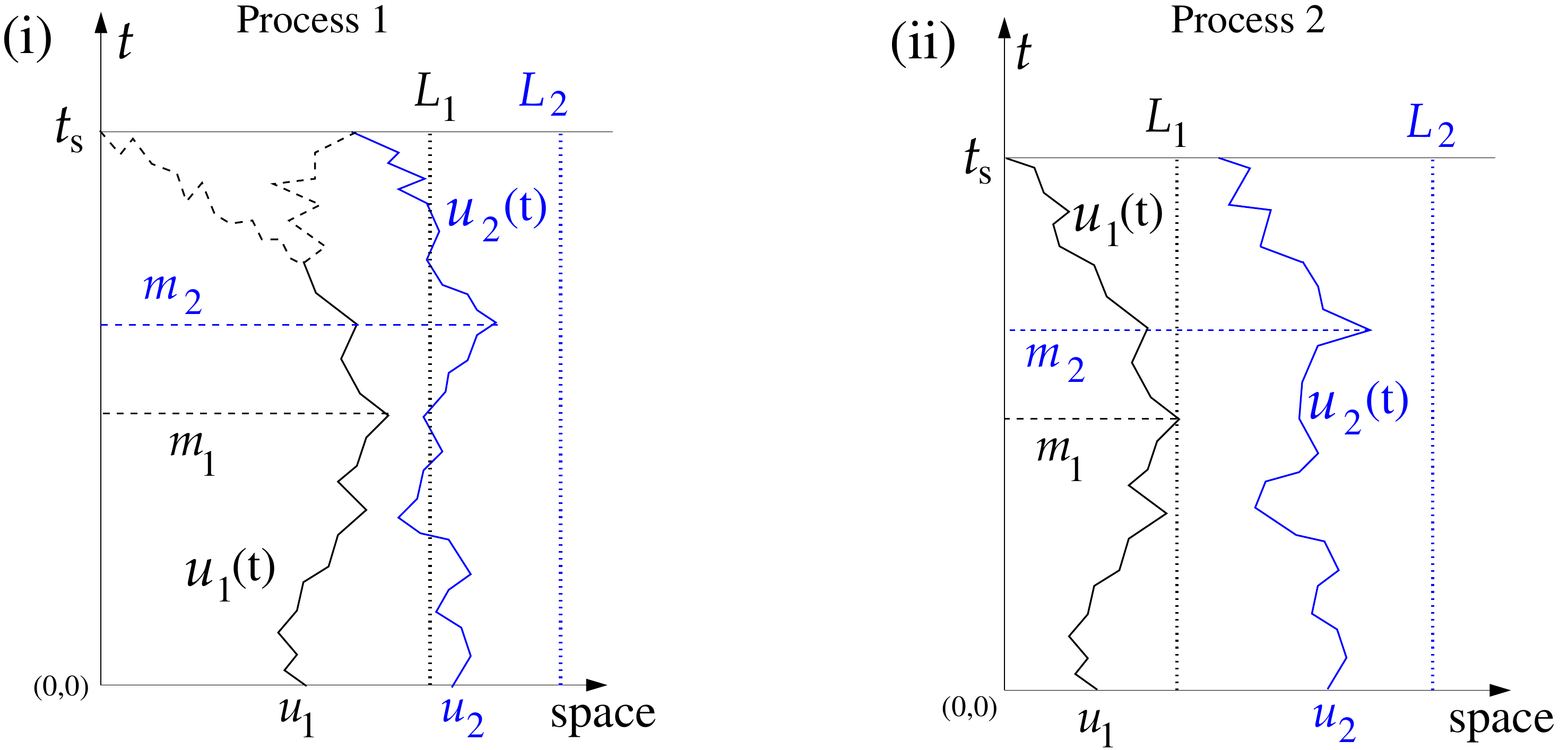}
\caption{(Color online) Schematic representation of the process 1 and process 2 with $N=2$ particles: $u_1(t)$ represents the trajectory of 
the 1st particle (the leftmost one) and 
$u_2(t)$ represents the trajectory of the 2nd particle (the rightmost one).}
\label{fig1}
\end{figure}

This is precisely the question which we address in this article, by considering non-intersecting Brownian motions, 
which is one of the simplest -- though non trivial -- interacting particles system. 
More precisely, we consider  $N$ Brownian walkers moving on a line with position $u_i(t)$ 
at time $t$ for $i=1,2,\dots,N$. 
They evolve with time according to the Langevin equations
\begin{equation}
 \frac{d}{dt}u_i(t)=\eta_i(t),~~\text{with}~~\langle \eta_i(t)\rangle=0~~\text{and}~~
 \langle \eta_i(t)\eta_j(t') \rangle =2D_i\delta_{ij}\delta(t-t') \;, \label{lang}
\end{equation}
where $D_i$ is the diffusion constant of the $i\text{th}$ particle and $\eta_i$'s are independent Gaussian white noises.
The initial positions of these particles are $u_i(0)=u_i$ such that $0 < u_1 < u_{2} <\dots < u_N$. 
The process gets stopped at a random time $t=t_s$ when a specific event occurs. In this paper we consider two different 
mechanisms of stopping event called ``process 1'' [see Fig. \ref{fig1} (i)] and ``process 2'' [see Fig.~\ref{fig1} (ii)]:
\begin{itemize}
\item{In ``process 1'', we consider 
the evolution of the $N$ 
Brownian walkers till time $t_s$ when  
either the first particle crosses the origin for the first time before any two walkers meet each other {\it or} 
any two particles meet for the first time before the first particle crosses the origin [see Fig. \ref{fig1} (i)].}
\item{In ``process 2'', $t_s$ is the time when 
the first particle crosses the origin for the first time before any two walkers meet  [see Fig.~\ref{fig1} (ii)].}
\end{itemize}
In both cases, the trajectories of the particles are non-intersecting. In the physics literature, such non-intersecting 
Brownian motions are called ``vicious walkers'' 
\cite{deGennes68,Fisher84} and have been recently studied in various contexts~\cite{Schehr08,Kobayashi,Nadal09,Forrester08,Izumi11,Rambeau11,Issac}.

Let $p_i(m|{\bf{u}})dm$, with $i = 1, 2, \cdots, N$, denote the probability that $m_i\in [m,m+dm]$, where  $m_i=\max_{0 \leq t \leq t_s} u_i(t)$ is the 
maximal distance travelled by the $i$th walker till the stopping time $t_s$. Here we mainly focus on the PDFs  $p_1(m|{\bf{u}})$ and $p_N(m|{\bf{u}})$ because 
$m_1$ and $m_N$ provide characterization of certain geometrical properties of the Brownian walker trajectories. 
For instance, one may think about $m_1$ as an estimate of the common region visited by all the $N$ walkers till the process ``stops'' 
(given that all the particles initially started very close 
to the origin). Similarly, $m_N$ characterizes the number of distinct sites visited by the $N$ walkers till $t_s$. Recently we have studied the distributions of the number of distinct sites and common sites visited by $N$ independent walkers over a fixed time interval 
$[0,t]$~\cite{Anupam13}. Our initial motivation was to generalize this case to interacting walkers over a fixed time interval. 
But it is a harder problem to solve. However we show in this paper that the problem with a
``stopping time'' is solvable even in the presence of interactions. It is also interesting to note that introducing an extra random 
variable namely the ``stopping time'' $t_s$ renders the problem analytically tractable.

Before presenting the details of our calculations, it is useful to give a summary of our results. 
We first study the $N=2$ particle problem because it is fully solvable even when the diffusion constants of the two particles 
are different \emph{i.e} $D_1 \neq D_2$ and also because the basic concepts are easy to present in this case. 
Solving a backward Fokker-Planck (BFP) equation we are able to find the full distributions $p_{2}(m|u_1,u_2)$ and $p_1(m|u_1,u_2)$ corresponding to 
the maximal displacements $m_2$ and $m_1$ of the right and left particle, respectively (see Fig. \ref{fig1}).
We show explicitly for both processes 1 and 2 that the PDFs $p_{2}(m|u_1,u_2)$ and $p_1(m|u_1,u_2)$ 
have power law tails valid for $m\gg u_2$, as
\begin{eqnarray}
 p_1(m|u_1,u_2) &\simeq& \dfrac{\mathcal{A}_1(u_1,u_2,D_1,D_2)}{m^{\nu_1}}\;, \label{2part-dist-intro1} \\
 p_2(m|u_1,u_2) &\simeq& \dfrac{\mathcal{A}_2(u_1,u_2,D_1,D_2)}{m^{\nu_2}}, \label{2part-dist-intro2} \\
 \text{with~exponents}~~~~~~~~\nu_1=\nu_2&=&\frac{3\pi - 2\arctan\left( \sqrt{\frac{D_1}{D_2}}\right)}
{\pi - 2\arctan\left( \sqrt{\frac{D_1}{D_2}}\right)} \;.\label{exponent-2part}
\end{eqnarray}
The functions $\mathcal{A}_i(u_1,u_2,D_1,D_2)$ are the amplitudes associated to the algebraic tails of the PDF 
$p_i(m|u_1,u_2)$ with $i=1,2$. While these amplitudes differ from process 1 to process 2, the exponents
$\nu_1=\nu_2$ (for the right and left walkers) are process independent. The amplitudes depend explicitly 
on the initial positions $u_1,~u_2$ as well as on the diffusion constants $D_1,~D_2$. 
Explicit expressions of $\mathcal{A}_i(u_1,u_2,D_1,D_2)$ for both processes 1 and 2 are given in Eqs.~(\ref{AaP11}) to (\ref{AaP22}).

Next we consider the general $N$-particle problem. In this case, based on the results for the non-interacting case [Eq. (\ref{non_int})] as well as 
on the results of the $N=2$-particle problem, one generally expects that the PDF $p_i(m|{\bf{u}})$ of the maximal 
distance $m_i$ of the $i^{\text{th}}$ particle till the stopping time $t_s$, has an algebraic tail:
\begin{equation}
 p_i(m|{\bf{u}}) \simeq \mathcal{A}_i({\bf{u},D})~\frac{1}{m^{\nu_i}},\;\;m\gg u_N,\;\;i=1,2, \cdots N \;.\label{pl}
\end{equation}
The exponents $\nu_i$'s and the amplitudes $\mathcal{A}_i$'s are, in general, different for the two processes for $N>2$ 
(note that for $N=2$, while the exponents are same, the amplitudes are different). They also 
depend explicitly on the number of particles $N$ and 
on the diffusion constants ${\bf{D}}=(D_1,D_2,\cdots,D_N)$. Proving the result in Eq. (\ref{pl}) for any $i=1,2,...,N$ and general $N$ 
is a hard task. However, one can make some progress for $i=N$ \emph{i.e} for the maximal distance $m_N$ travelled by the rightmost walker. When the walkers are 
identical \emph{i.e.} when they have identical diffusion constants $D_1=D_2=...=D_N=D$, we estimate the tail of the PDF $p_N(m|{\bf{u}})$ using a 
heuristic scaling argument based on the distribution $f_N(t_s|{\bf u})$ of the ``stopping time'' $t_s$. This argument, for both processes 1 and 2, yields :
\begin{equation}
 \nu_N = N^2 +1,~~\text{when}~~D_1=D_2=...=D_N=D. \label{nu_N}
\end{equation}
We also obtain an explicit expression of the prefactor $\mathcal{A}_N({\bf u}, D_1 = D, \cdots, D_N=D)$ in Eq. (\ref{pl}). We observe that for identical diffusion constants this prefactor does not depend on $D$ explicitly. Hence suppressing $D$ from the argument, 
we denote $\mathcal{A}_N({\bf u}, D_1=D, \cdots, D_N = D) = \mathcal{A}_N({\bf u})$ and show that it is given by
\begin{eqnarray}
&&\mathcal{A}_N({\bf u}) \approx N^2 B_N \mathcal{F}({\bf u}) \;,~~\text{with} \;~\mathcal{F}({\bf u}) =\frac{\mathcal{Y}_N({\bf u})}{S_N({\bf u})},\label{result-p_N} \\
&& \text{where}\,~\mathcal{Y}_N({\bf u}) = \prod_{i=1}^{N}u_i\;\prod_{1\leq i<j \leq N}(u_j^2-u_i^2), \label{mathcalY_intro}
\end{eqnarray}
and $S_N({\bf u})$ is an exit probability whose value is $1$ for process 1 and smaller than $1$ for process 2 [given in Eq. (\ref{S_N-explicit})].
We also present a formal exact expression of the $N$ dependent constant $B_N$, which for large $N$, is shown to grow asymptotically as 
\begin{equation}
 B_N \approx \text{exp}\left[\frac{N^2}{2} \log N + o(\log N) \right] \;, \label{B_N-intro}
\end{equation}
where $o(\log N)$ represents terms smaller than $\log N$. This large $N$ asymptotic form of $B_N$ should be compared with the corresponding behavior $b_N \approx \text{exp}\left[\frac{N}{2}\log (\log N)\right]$ 
in the non-interacting case in Eq.~(\ref{non_int}).

The paper is organized as follows. In section \ref{section3} we consider the two walkers problem where we evaluate the PDFs $p_1(m|u_1,u_2)$ and $p_{2}(m|u_1,u_2)$ corresponding 
to $m_1$ and $m_2$ respectively. In this section we solve a BFP equation, which under the ``stopping time'' framework becomes a Laplace's equation. From the 
solution of the BFP equation we find the distributions of the individual maximal distances of the first and second particles. 
In section \ref{multipart} we consider the general $N$-particle problem. This section is divided into two subsections. In the first subsection \ref{argument}, we give a heuristic scaling 
argument based on the distribution of the ``stopping time'' $t_s$, to find the exponent $\nu_N$ of the 
power law tail of the PDF $p_N(m|{\bf u})$ corresponding to the global maximum $m_N$. In the next subsection \ref{morethan2}, 
we present a more rigorous calculation based on $N$-particle Green's function to establish the power law obtained in the previous section \ref{argument}. This calculation 
also provides exact expressions for the amplitudes associated to the tail of $p_N(m|{\bf u})$. Some technical details have been left in Appendices \ref{delta_N}, \ref{expreuneqD}, \ref{Psi-function} and \ref{B_N-lrg-N}.


\section{Two walkers problem ($N=2$): exact solution}\label{section3}
\noindent 
Let us consider the motion of two non-identical Brownian walkers $u_1(t)$ and $u_2(t)$ given by 
\begin{eqnarray}
\dot u_i(t) &=& \eta_i(t) \;, \text{with}~\langle \eta_i(t)\rangle=0,~\text{for}~i=1,2 \;, \label{lang-2-part} \\
\text{and}~&&\langle \eta_1(t)\eta_1(t') \rangle =2D_1\delta(t-t'), \nonumber \\
&&\langle \eta_2(t)\eta_2(t') \rangle =2D_2\delta(t-t'), \nonumber \\
&&\langle \eta_1(t)\eta_2(t') \rangle =0,
\end{eqnarray}
where $D_1$ and $D_2$ are the diffusion constants of the first (left) and second (right) particle respectively.
To compute the PDFs of the individual maximum displacements $m_1$ and $m_2$, respectively, of the first and second particle, 
we start by defining the joint cumulative distribution function
\begin{equation}
\mathcal{Q}({\bf{L}}|{\bf{u}})\equiv \mathcal{Q}(L_1,L_2|u_1,u_2)=\text{Prob.}[m_1\leq L_1,~m_2\leq L_2|~0<u_1<L_1;~u_1<u_2<L_{2}],\label{jntQ}
\end{equation}
given that the initial positional order is 
maintained till $t_s$ and ${\bf{L}}=(L_1,L_2)$. The marginal cumulative distribution 
$\mathcal{Q}_1(L|u_1,u_2)=\text{Prob.}[m_1\leq L|~0<u_1<L;~u_1<u_2<\infty]$ is obtained by taking the limits $L_1 \to L$ and $L_2 \to \infty$ in
$\mathcal{Q}({\bf{L}}|{\bf{u}})$ whereas the marginal cumulative distribution $\mathcal{Q}_2(L|u_1,u_2)=\text{Prob.}[m_2\leq L|~0<u_1<L;~u_1<u_2<L]$ 
is obtained by taking the limit $L_1 \to L_2=L$ of
$\mathcal{Q}({\bf{L}}|{\bf{u}})$. 

 To find $\mathcal{Q}({\bf{L}}|{\bf{u}})$ we consider a different problem. We consider the first exit problem of a 
 single Brownian walker ${\bf u}(t) = (u_1(t),u_2(t))$ moving in 
 two dimensions inside the region $\mathcal{W}=$
\begin{tikzpicture}
\node [draw,trapezium,trapezium left angle=70,trapezium right angle=-90,minimum height=0.1cm] {};
\end{tikzpicture}
$OBCD$ described in Fig.~\ref{fig2} (i). We are interested in the probability with which the 2d-walker exits from 
$\mathcal{W}=$
 \begin{tikzpicture}
\node [draw,trapezium,trapezium left angle=70,trapezium right angle=-90,minimum height=0.1cm] {};
\end{tikzpicture}
$OBCD$ through specific boundaries for the first time. 
We denote this first exit probability by $F({\bf{u}};{\bf{L}})$ for both processes 1 and 2. 

For process 1, the exit probability $F({\bf{u}};{\bf{L}})$ represents the probability that 
the 2d-walker, starting from position $(u_1,u_2)$, 
exits from the region $\mathcal{W}$ through boundary $OD$ or $OB$ for the first time. When the 2d-walker exits through $OB$, it corresponds, in the original 
two-particle picture (Fig. \ref{fig1}), to the first (left) particle meeting the second (right) particle before it hits the origin for the first time at $t=t_s$ 
while keeping $m_1 \leq L_1$ and $m_2\leq L_2$ over $[0,t_s]$. 
In contrast, first exit of the 2d-walker through $OD$ corresponds to the first 
particle hitting the origin before meeting the second particle for the first time at $t=t_s$ while maintaining $m_1 \leq L_1$ and $m_2\leq L_2$ over $[0,t_s]$.
On the other hand, for process 2 the function $F({\bf{u}};{\bf{L}})$ represents the probability that the 
2d-walker exits from the region $\mathcal{W}$ only through boundary $OD$ for the first time. This exit event, in the two-particle picture, corresponds to the first 
particle hitting the origin for the first time before the two particles meet each other 
while keeping $m_1 \leq L_1$ and $m_2\leq L_2$. In the limit $L_1 \to \infty$ and $L_2 \to \infty$,  
we get the ultimate exit probability 
\begin{equation}
 S_2(u_1,u_2)=\lim \limits_{L_1 \to \infty}\lim \limits_{L_2\to \infty}
 F({\bf{u}};{\bf{L}}), \label{Ult-Survivl}
\end{equation}
which, for process 1, represents the the probability that the first particle hits either the origin or 
the second particle ultimately. Of course this occurs with probability $S_2(u_1,u_2)=1$ in this case. 
On the other hand, for process 2, $S_2(u_1,u_2)$ represents the probability that the first particle 
hits the origin for the first time before it collides with the second particle. 
This exit probability $S_2(u_1,u_2)$, in case of process 2, is precisely the survival probability of a lamb in the so-called ``lamb-lion'' problem where
it is being chased by a single diffusing lion in the presence of a refuge. If we identify the first particle as the lamb, the second particle as the  
lion and the origin as the refuge \cite{Rednerbook,Gabel12} then $S_2(u_1,u_2)$ is the probability that the lamb survives (i.e. reaches the refuge)
before being caught by the lion. This probability is smaller than one since there is a finite probability that the lion catches the lamb (i.e. the second particle hits 
the first particle before the later hits the origin). In particular for process 2, one can show that \cite{Gabel12}
\begin{eqnarray}
\label{Q}
 S_2(u_1,u_2) = 
 1- \arctan \left(\dfrac{u_1}{u_2}\sqrt{\dfrac{D_2}{D_1}}\right)\left[ \arctan \left(\sqrt{\dfrac{D_2}{D_1}}\right)\right]^{-1},
\label{exit-ult-2part}
 \end{eqnarray}
which for $D_1=D_2=D$ becomes independent of $D$ and is given by $S_2(u_1,u_2)= 1-\frac{4}{\pi}\arctan (u_1/u_2)$.

The quantity $S_2(u_1,u_2)$ has nice interpretations in terms of the trajectories of the two walkers. 
It represents the volume of a set of trajectories which contains 
all pairs of such trajectories which, starting from positions $(u_1,u_2)$, stay non-intersecting till $t_s$, whereas 
the quantity $F({\bf{u}};{\bf{L}})$ represents the volume of a subset, which 
contains such pairs of non-intersecting trajectories that are constrained by $m_1 \leq L_1$ and 
$m_2 \leq L_2$. Hence the ratio $\frac{F({\bf{u}};{\bf{L}})}{S_2(u_1,u_2)}$ gives the fraction of such pairs of vicious 
trajectories which have $m_1 \leq L_1$ and $m_2 \leq L_2$. 
This fraction precisely represents the cumulative probability $\mathcal{Q}({\bf{L}}|{\bf{u}})$ defined 
in Eq. (\ref{jntQ}). Hence, if we know the exit probability $F({\bf{u}};{\bf{L}})$, the cumulative probability $\mathcal{Q}({\bf{L}}|{\bf{u}})$ 
is obtained from 
\begin{equation}
 \mathcal{Q}({\bf{L}}|{\bf{u}})= \frac{F({\bf{u}};{\bf{L}})}{S_2(u_1,u_2)}, \;\;\;
 \text{where}~~S_2(u_1,u_2)=\lim \limits_{L_1 \to \infty}\lim \limits_{L_2\to \infty}
 F({\bf{u}};{\bf{L}}).\label{P0}
\end{equation}
The next question is then how to compute this exit probability $F({\bf{u}};{\bf{L}})$ in Eq. (\ref{P0}). 
In the next subsection we show that the probability $F({\bf{u}};{\bf{L}})$ satisfies a Laplace's equation which we solve 
with boundary conditions specified for both process 1 and process 2. 

\subsection{Backward Fokker-Planck equation for $F({\bf{u}};{\bf{L}})$}
\label{BkFP}
\noindent 
A powerful tool to study the PDF of first passage times, like $t_s$ in our problem [see Fig. (\ref{fig1})], is the backward 
Fokker-Planck equation \cite{Rednerbook,Bray13}. Here we are actually dealing with functionals of 
$t_s$, $m_i=\max \{ x_i(t), 0\leq t \leq t_s\}$.
For such functional also, it is possible to use an approach based on BFP equation (see Ref. \cite{satyacurrsci} for a review). 
Here we write down a BFP equation for the quantity $F({\bf{u}};{\bf{L}})$ treating 
the initial coordinates $u_i$ as independent variables. To do this, we consider trajectories $u_1(t)$ and $u_2(t)$ of the two Brownian 
particles over the interval $[0,t_s]$, which evolve according to the Langevin Eqs. (\ref{lang}). We first split the time interval $[0,t_s]$ into two 
parts: 
$[0,\Delta t]$ and $[\Delta t, t_s]$. In the first infinitesimal time window $[0,\Delta t]$ the two Brownian particles will move from their initial 
positions $\{u_1,u_2\}$ to new positions $\{u_1+\Delta u_1,u_2+\Delta u_2\}$, where 
\begin{equation}
 \Delta u_1 = \int_0^{\Delta t} \eta_1(t')dt'~\text{and}~\Delta u_2 = \int_0^{\Delta t} \eta_2(t')dt' \;. \label{DX}
\end{equation}
These two new positions are considered as ``new" initial positions of the two Brownian particles, respectively, for the evolution in the subsequent 
time interval $[\Delta t, t_s]$. Since the evolution of the positions of the particles are Markovian, we have 
\begin{equation}
 F({\bf{u}};{\bf{L}})= \Big \langle F({\bf{u}}+\Delta {\bf{u}};{{\bf L}})\Big \rangle_{\Delta {{\bf u}}} . \label{F-markov}
\end{equation}
By Taylor expanding the right hand side of the above equation in $\Delta u_1, \Delta u_2$ we have 
\begin{equation}
F({\bf{u}};{\bf{L}}) =  F({\bf{u}};{\bf{L}}) + \sum_{i=1}^2 \left[ \frac{\partial F}{\partial u_i} \left \langle \Delta u_i \right \rangle
+\frac{1}{2}\frac{\partial^2 F}{\partial u_i^2} \left \langle \Delta u_i^2 \right \rangle \right] 
+ \frac{\partial^2 F}{\partial u_1\partial u_2} \left \langle \Delta u_1 \Delta u_2 \right \rangle + \dots
\label{taylor}
\end{equation}
From the Langevin Eqs. (\ref{lang}) one can easily show that 
\begin{equation}
 \left \langle \Delta u_i \right \rangle = 0 \; ,\;
 \left \langle \Delta u_i \Delta u_j \right \rangle = 2\delta_{ij}D_i\Delta t~~\text{for}~~~i=1,2 \;. 
\end{equation}
Using these relations in Eq. (\ref{taylor}) and keeping only terms of $\mathcal{O}(\Delta t)$ we obtain the following partial 
differential equation
\begin{equation}
  D_1\frac{\partial^2 F}{\partial u_1^2}+ D_2\frac{\partial^2 F}{\partial u_2^2} =0 \label{FP1} \;,
\end{equation}
with boundary conditions (BCs) determined by the stopping rules, which are thus different for process 1 and process 2. The Eq.~(\ref{FP1}) 
is valid over the region 
$\mathcal{W}=\{0\leq u_1 \leq L_1; u_1\leq u_2 \leq L_{2}\}$.
\begin{figure}[t]
\centering
\includegraphics[scale=0.35]{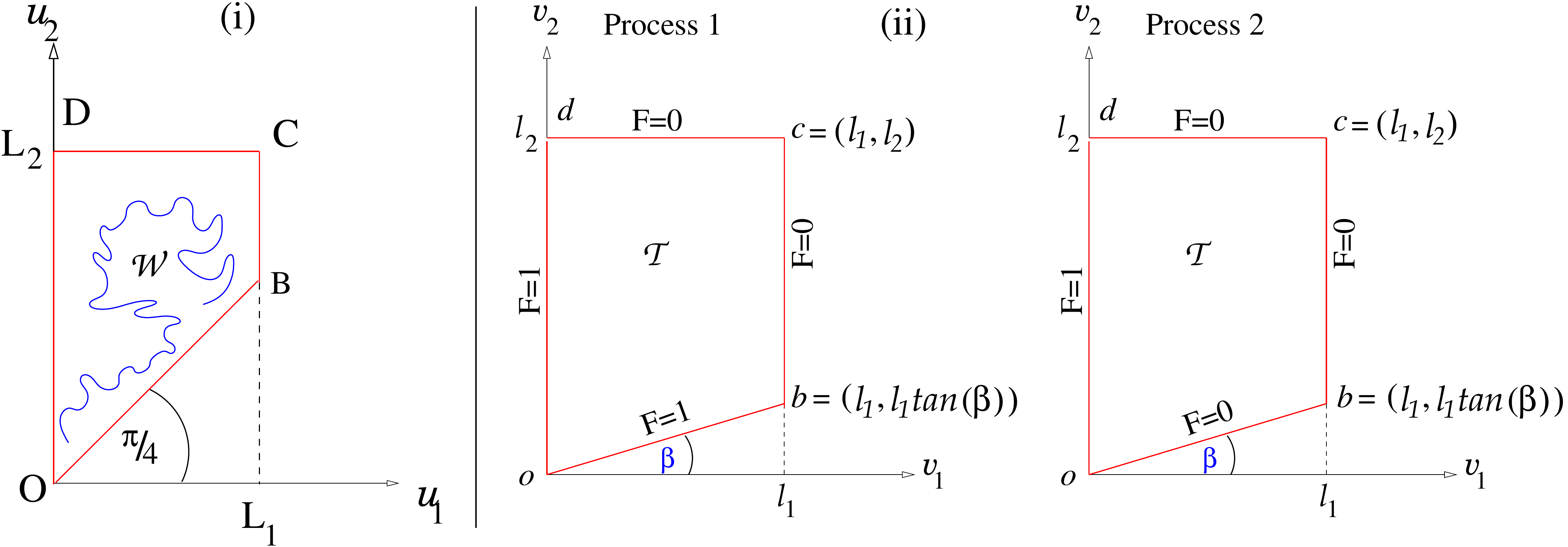}
\caption{(Color online) (i) Motion of a 2d-walker inside the region $\mathcal{W}$, 
(ii) Boundaries and boundary conditions associated to the Laplace's equation in~(\ref{laplace1}) 
for process 1 and process 2. Note that $l_i = L_i\sqrt{D_i}$ and $\tan(\beta)=\sqrt{D_1/D_2}$.}
\label{fig2}
\end{figure}
It is convenient to perform the following rescaling
\begin{eqnarray}
 v_i=\frac{u_i}{\sqrt{D_i}} \;,~~
 l_i=\frac{L_i}{\sqrt{D_i}}~,\label{trans}
\end{eqnarray}
which transforms the trapezium OBCD in the $(u_1,u_2)$ plane to the trapezium $obcd$ in the $(v_1,v_2)$ plane where $\tan(\beta)=\sqrt{D_1/D_2}$.
Under this transformation the Fokker-Planck equation in (\ref{FP1}) becomes the Laplace's equation
\begin{equation}
\frac{\partial^2 F}{\partial v_1^2}+ \frac{\partial^2 F}{\partial v_2^2}=0, \label{laplace1}
\end{equation}
which holds over the region $\mathcal{T}=$
\begin{tikzpicture}
\node [draw,trapezium,trapezium left angle=70,trapezium right angle=-90,minimum height=0.1cm] {};
\end{tikzpicture}
$obcd=\{0\leq v_1 \leq l_1; v_1\tan(\beta) \leq v_2 \leq l_{2}\}$ 
[see Fig.~\ref{fig2} (ii)] with appropriate BCs. We give the BCs in Table \ref{BCs}, which 
can be understood from the following arguments:
\begin{table}[hh]
\centering
\begin{tabular}{|c| c|c |}
\hline
\multicolumn{3}{|c|}{ Boundary conditions with $\tan(\beta)=\sqrt{\frac{D_1}{D_2}}$}\\
\hline
 Boundary & process 1 &  process 2 \\ 
\hline
$v_1=0$ ~~~~~~~~~~~[{\it{od}}] & $F(v_1=0,v_2;l_1,l_2)=1$ & $F(v_1=0,v_2;l_1,l_2)=1$ \\
\hline
$v_2=\tan{(\beta)}v_1$ ~[{\it{ob}}] & $F(v_1,~v_2=\tan(\beta)v_1;l_1,l_2)=1$ &  $F(v_1,~v_2=\tan(\beta)v_1;l_1,l_2)=0$\\
\hline
$v_1=l_1$ ~~~~~~~~~~[{\it{bc}}]& $F(v_1=l_1,v_2;l_1,l_2)=0$ & $F(v_1=l_1,v_2;l_1,l_2)=0$ \\
\hline
$v_2=l_2$ ~~~~~~~~~~[{\it{cd}}]& $F(v_1,v_2=l_2;l_1,l_2)=0$ & $F(v_1,v_2=l_2;l_1,l_2)=0$ \\
\hline
\end{tabular}
\caption{Table of boundary conditions associated to the Laplace's equation in~(\ref{laplace1}) for process 1 and process~2.}
\label{BCs}
\end{table}
\begin{itemize}
 \item 
BC on the segment $[od]$ ($v_1=0)$: If the first particle starts with $u_1=0$ and $u_1\leq u_2 \leq L_2$ (\emph{i.e.} $v_1=0$ and 
$0 \leq v_2 \leq l_2$), 
then the first particle immediately crosses the origin, which implies $t_s=0$, for both process 1 and process 2. Clearly then the maximal 
displacement of the two particles remains $m_1=0 < L_1$ and $m_2=u_2 \leq L_2$ implying the BC $F(v_1=0,v_2;l_1,l_2)=1$ on $[od]$ [see 
Fig.~\ref{fig2} (ii)] for both processes 1 and 2. 

\item
BC on the segment $[ob]$ [$v_2=\tan(\beta)v_1$]: If both particles start from the same position {\it i.e.} $u_1=u_2<L_1 < L_2$, then they 
immediately collide implying $t_s=0$ and hence, for process 1, $F(v_1,v_2=\tan(\beta)~v_1;l_1,l_2)=1$. On the other hand, process 2 excludes 
the possibility of any collision between the two particles even at time $t=t_s$. This implies 
$F(v_1,v_2=\tan(\beta)~v_1;l_1,l_2)=0$
instead of $1$ [see Fig.~\ref{fig2} (ii)]. 

\item
BC on the segment $[bc]$ ($v_1=l_1$): When the initial positions are say $u_1=L_1$ and $L_1\leq u_2 \leq L_2$ (\emph{i.e.} 
$v_1=l_1$ and $l_1\leq v_2 \leq l_2$), then clearly $m_1=L_1$ at $t=0$ and it will definitely become larger than $L_1$ in the next 
subsequent instant. Hence the BC on the segment $[bc]$ is $F(v_1=l_1,v_2;l_1,l_2)=0$ [see Fig.~\ref{fig2} (ii)] for both processes 1 and 2. 

\item 
BC on the segment $[cd]$ ($v_2=l_2$): When the second particle starts from its initial position $u_2=L_2$ ({\it i.e.} $v_2=l_2$) then $m_2=L_2$ right at the beginning 
and $m_2$ will definitely become larger than $L_2$ in the next 
subsequent instant implying $F(v_1,v_2=l_2;l_1,l_2)=0$ on the segment $[cd]$ [see Fig.~\ref{fig2} (ii)] for both processes 1 and 2. 
\end{itemize}

To summarize, we finally have to solve a Laplace's equation in 
(\ref{laplace1}), which holds inside the polygon ${\mathcal{T}}$ in the plane $(v_1,v_2)$ shown in Fig. \ref{fig2} 
with BCs specified in Table \ref{BCs} for both process 1 and process 2. 

\subsection{Solution of the Laplace's equation via conformal mapping}
\label{confmap}
\begin{figure}[h]
\centering
\includegraphics[scale=0.35]{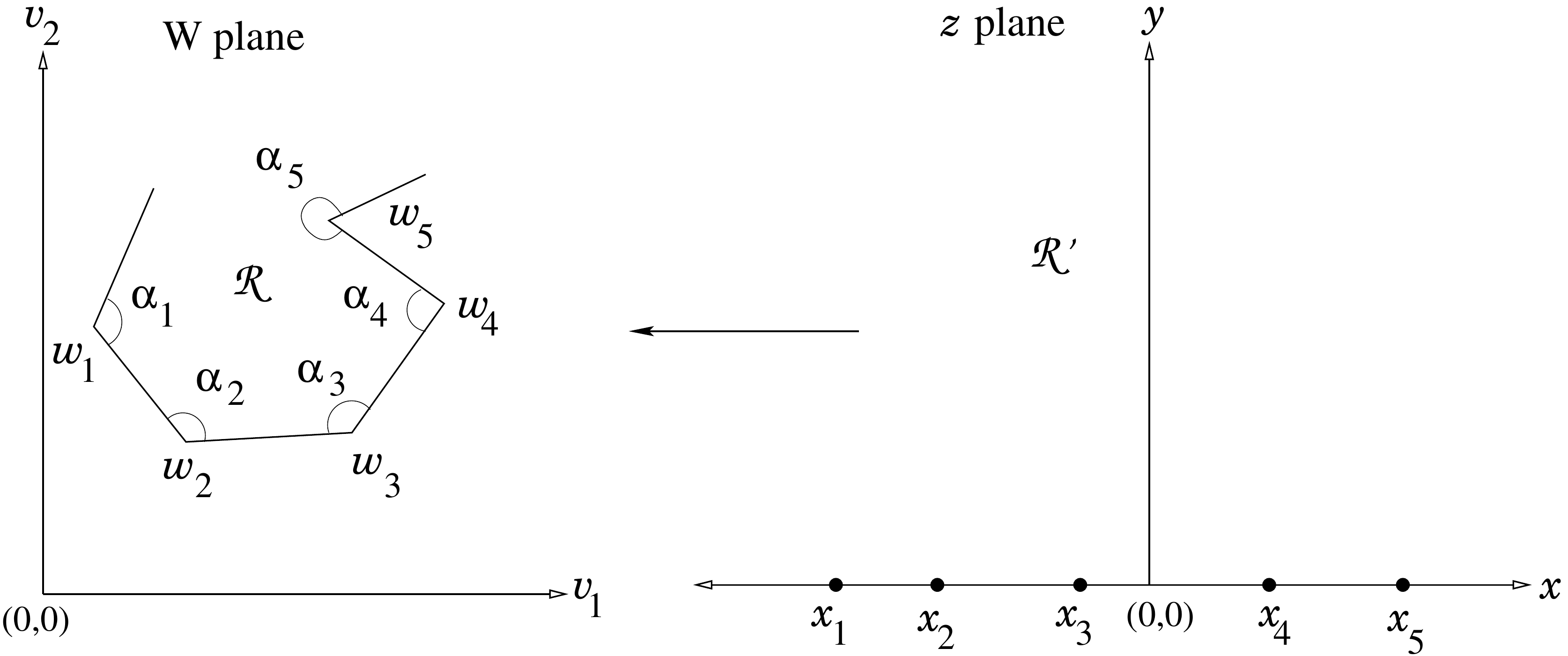}
\caption{(Color online) Schematic representation of Schwarz-Christoffel transformation $W(z)$ in Eq. (\ref{SCtrans}), such that $w_i = W(x_i)$.}
\label{fig3}
\end{figure}

\begin{figure}[h]
\centering
\includegraphics[scale=0.35]{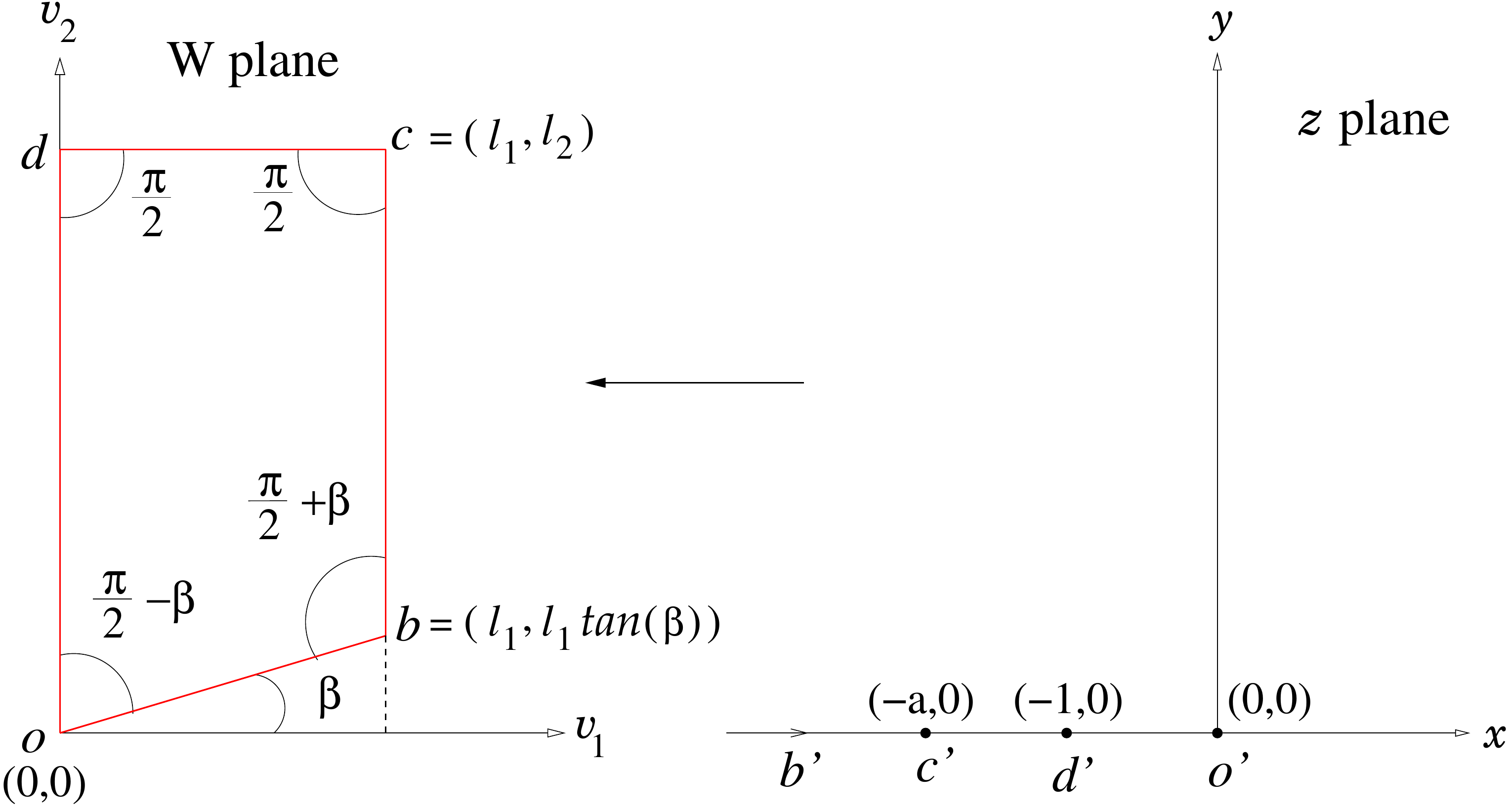}
\caption{(Color online) Schematic representation of Schwarz-Christoffel transformation in Eq. (\ref{SCtrans1}).}
\label{fig4}
\end{figure}

\noindent 
Solving the Laplace's equation in (\ref{laplace1}) for any given BC is not {\it a priori} an easy task. However using a conformal transformation of the 
variables, one can transform boundaries of the domain ${\cal T}$ to a much simpler geometry, while leaving the Laplace's equation itself invariant. 
Following Ref. \cite{SatyaBray2010}, we here use the Schwarz-Christoffel (S-C) transformation which operates as follows:
For a polygon $\mathcal{P}$ (see Fig. \ref{fig3}) in the $W$ plane having $n$ vertices
$\{w_1,w_2,\dots, w_n\}$ with corresponding interior angles $\{\alpha_1,\alpha_2,\dots,\alpha_n\}$, there exists a transformation $W=W(z)$ from 
complex $z$-plane to $W$ plane such that the upper half $\mathcal{R}'$ of the $z$-plane gets mapped onto the interior region $\mathcal{R}$ of the 
polygon in the $W$ plane. 
Under this transformation $W=W(z)$, the real axis in the $z$-plane gets mapped onto the boundary of the polygon $\mathcal{P}$ with the $n$ 
vertices $\{w_1,w_2,\dots, w_n\}$ being 
images of the $n$ specific points $\{x_1,x_2,\dots, x_n\}$ on the real axis. As a result, solving the Laplace's equation with complicated boundaries 
reduces to finding the electrostatic potential on the upper half of the complex $z$-plane when the 
potential is given on the real axis: the electrostatic potential can then be obtained explicitly from the Poisson's integral formula. 
The S-C transformation reads as \cite{Schaums}
\begin{equation}
 W(z)=B_0~\int_0^z(t-x_1)^{\frac{\alpha_1}{\pi}-1}(t-x_2)^{\frac{\alpha_2}{\pi}-1}\dots (t-x_n)^{\frac{\alpha_n}{\pi}-1}~dt + C_0, \label{SCtrans}
\end{equation}
where $B_0$ and $C_0$ are arbitrary constants. 
It is convenient to choose one point, say $x_n$, at $-\infty$, such that the last factor $(t-x_n)^{\frac{\alpha_n}{\pi}-1}$ present in the integrand of 
Eq. (\ref{SCtrans}) is absent. 
In our problem, we have a trapezium $obcd$ as shown in Fig.~\ref{fig4} for both processes 1 and 2. We chose a point $b'$ on the real line of the $z$-plane at $-\infty$, which 
corresponds to the image of vertex $b$ on the $W$ plane (see Fig. \ref{fig4}). Moreover, let us consider that the points 
$c',d',o'$ on the real line with coordinates $x=-a,~x=-1$ and $x=0$ are mapped onto the vertices $c,d,o$ (see Fig. \ref{fig4}) 
under the transformation $W(z)$. 
One thus has
\begin{equation}
W(z)=B_0~\int_0^z t^{-(\frac{\beta}{\pi}+\frac{1}{2})}(t+1)^{-\frac{1}{2}}(t+a)^{-\frac{1}{2}}~dt + C_0,~~~\text{where}~~~
 \beta=\arctan \left( \sqrt{\frac{D_1}{D_2}} \right ),\label{SCtrans1}
\end{equation}
where $a$, $B_0$ and $C_0$ are unknown constants to be determined. 
Since in our case the origin is mapped onto itself 
under the transformation $W(z)$, i.e. $W(0)=0$, we have $C_0=0$. Hence, from Eq. (\ref{SCtrans1}) and Eq. (\ref{trans}) we have 
\begin{equation}
\frac{u_1}{\sqrt{D_1}} + i \frac{u_2}{\sqrt{D_2}}= W(z)= B_0\int_0^z \frac{dt}{t^{1-\theta}\sqrt{1+t}\sqrt{a+t}}~~~
\text{where}~~~\theta =\frac{1}{2}-\frac{1}{\pi}\arctan \left( \sqrt{\frac{D_1}{D_2}}\right ). \label{SCtrans2}
\end{equation}
Here we note that for $D_1=D_2$ the exponent $\theta=\frac{1}{4}$. 
The unknown constants $B_0$ and $a$ in Eq. (\ref{SCtrans2}) are determined as follows. The points 
$c' \equiv (-a,0)$ and $d' \equiv (-1,0)$ on the real axis get mapped onto 
the points $c \equiv (l_1=\frac{L_1}{\sqrt{D_1}},l_2=\frac{L_2}{\sqrt{D_2}})$ 
and $d \equiv (0, l_2)$ on the $W$ plane respectively, which implies
\begin{eqnarray}
 d=W(d') \implies~~i~l_1\alpha \tan(\beta) &=& B_0 \int_0^{-1} \frac{t^{\theta-1}}{\sqrt{1+t}\sqrt{a+t}}dt \label{B1}, \\
c=W(c') \implies~~ l_1(1+i~\alpha \tan(\beta)) &=& B_0 \int_0^{-a} \frac{t^{\theta-1}}{\sqrt{1+t}\sqrt{a+t}}dt \label{C1},
\end{eqnarray}
where the variable $\alpha$ is defined as
\begin{eqnarray}
\alpha={L_2}/{L_1}. \label{def_alpha}
\end{eqnarray}
Simplifying Eqs. (\ref{B1}) and (\ref{C1}) one obtains the following two expressions,  
\begin{eqnarray}
 &&~~~~~~~~~~~~~~~~~~~\frac{h_{\theta}(a)}{k_{\theta}(a)} =\frac{1}{\alpha\tan(\beta)}=\frac{L_1\sqrt{D_1}}{L_2\sqrt{D_2}}, \label{a} \\ 
 &&B_0= \frac{l_1\alpha\tan(\beta)}{k_{\theta}(a)}~e^{i\beta}=\frac{L_2}{k_{\theta}(a)\sqrt{D_2}}
 \text{exp}\left[ i \arctan \left( \sqrt{\frac{D_1}{D_2}} \right )\right],\label{B_0} 
 \end{eqnarray}
 \begin{equation}
  \text{with}~~h_{\theta}(a)=\int_1^a \frac{t^{\theta-1}}{\sqrt{t-1}\sqrt{a-t}}~dt ~~
 \text{and}~~~k_{\theta}(a)= \int_0^1 \frac{t^{\theta-1}}{\sqrt{1-t}\sqrt{a-t}}~dt \label{hka} \;,
 \end{equation}
 which determine the two constants $a$ and $B_0$.
 \begin{figure}[h]
\centering
\includegraphics[scale=0.4]{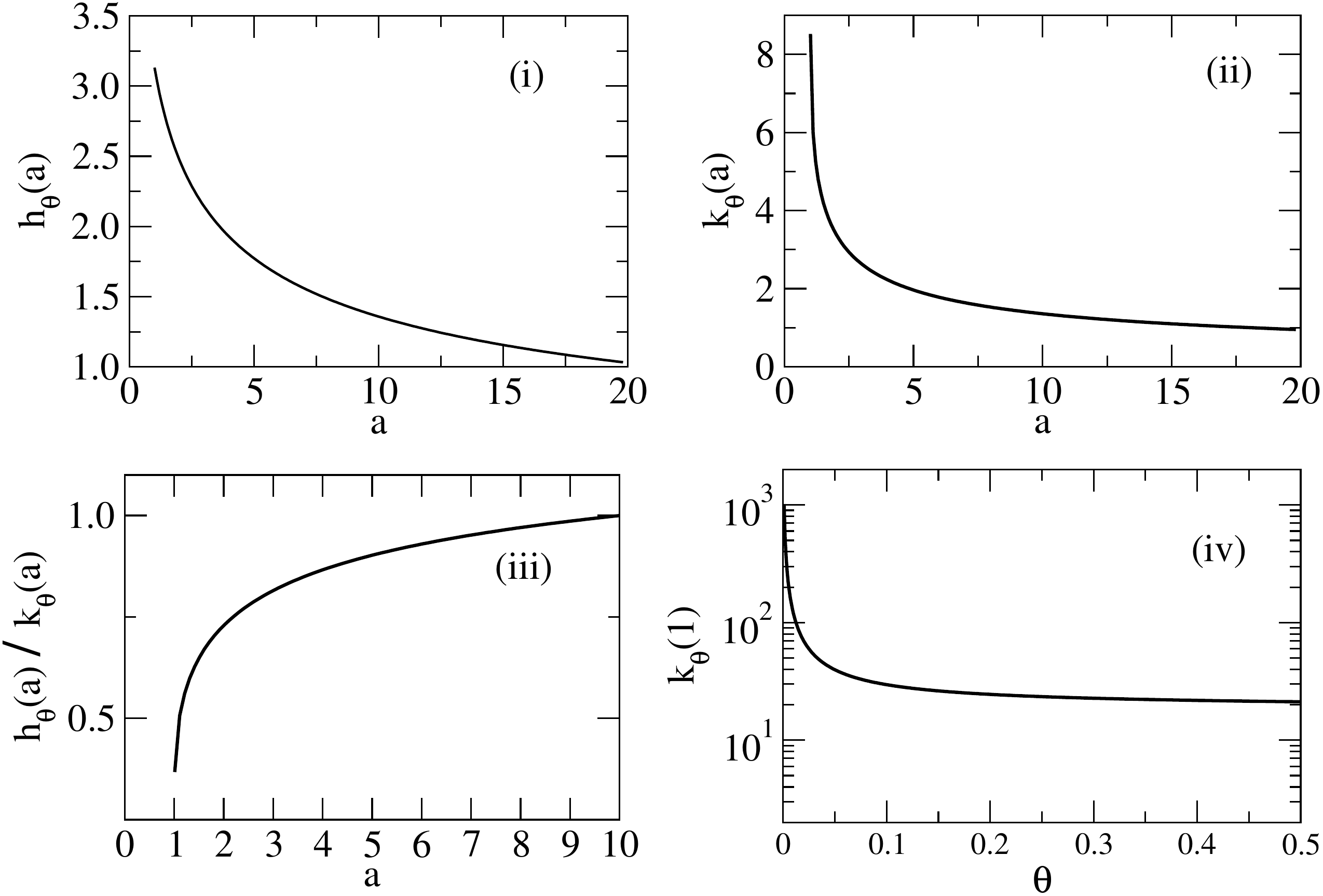}
\caption{(Color online) Properties of the integrals $h_{\theta}(a)$ and $k_{\theta}(a)$ given in Eq. (\ref{hka}). 
In figures (i), (ii) and (iii) we have used the value $\theta={1}/{3}$.}
\label{fig5}
\end{figure}
The solution of the Eq. (\ref{a}) gives the value of $a$ for given $\alpha$ and $\beta$ whereas using this solution for $a$ in Eq. (\ref{B_0}) 
we get $B_0$. 
When $\alpha \to 1$, \emph{i.e.} $L_1 \to L_2$, the vertices $c$ and $b$ of the trapezium $obcd$ approach to each other. This means that   
the point $c'$ on the real axis of the $z$-plane (Fig. \ref{fig4}) should approach $b'$ implying $a \to \infty$ as $L_1 \to L_2$. On the other hand, 
when $\alpha \to \infty$,  \emph{i.e.} $L_2 \gg L_1$, the point $c'$ should  
approach $d'$ implying $a \to 1$ in this limit. Hence we expect that the value of $a$ should lie in the interval $[1, + \infty)$ 
for $1<\alpha<\infty$ where both 
the integrals $h_{\theta}(a)$ and $k_{\theta}(a)$ are smooth real valued functions of $a$ for given $\theta$. 
In Fig. \ref{fig5} (i) and (ii) we show how the integrals  $h_{\theta}(a)$ and $k_{\theta}(a)$ behave as a function of $a$ for $\theta=\frac{1}{3}$.
Solving the Eq. (\ref{a}) we obtain the value of $a$ for given $\alpha$ and $\beta$ and using this value of $a$ in Eq. (\ref{B_0}) 
we can get $B_0$. In Fig. \ref{fig5bis} we plot $a$ as a function of $\alpha$ for $D_1=1.3$ and $D_2=1.5$ and see that $a$ 
diverges when $\alpha$ goes to $1$ 
whereas $a$ approaches $1$ when $\alpha \to \infty$ (as expected from the above arguments). 
Let us analyze the integrals 
$h_{\theta}(a)$ and $k_{\theta}(a)$ in detail to see how $a$ behaves as a function of $\alpha$ when $\alpha \to 1$ and 
$\alpha \to \infty$ separately. 

We first consider the case when $\alpha$ approaches $1$ from above {\it{i.e.}} $a \to \infty$. Expanding the 
two functions entering the left hand side (l.h.s.) of Eq. (\ref{a}) for large $a$ we get
\begin{eqnarray}
 h_{\theta}(a) &\simeq& \frac{\sqrt{\pi}\Gamma[\theta]}{\sqrt{a}\tan{\beta}}\left[\frac{1}{\Gamma[\frac{1}{2}+\theta]} 
 + \frac{\theta}{\Gamma[\frac{3}{2}+\theta]} \frac{1}{a} + \mathcal{O}(a^{-2}) \right], \label{hatinf} \\
 k_{\theta}(a) &\simeq & \frac{\sqrt{\pi}\Gamma[\theta]}{\sqrt{a}}\left[\frac{1}{\Gamma[\frac{1}{2}+\theta]} 
 + \left(\frac{\theta}{2\Gamma[\frac{1}{2}+\theta]} - \frac{\theta}{4\Gamma[\frac{3}{2}+\theta]}\right )\frac{1}{a} + \mathcal{O}(a^{-2}) \right],
 \label{katinf}
\end{eqnarray}
where $\Gamma[x]$ is the Gamma function.
\begin{figure}[ht]
\centering
\includegraphics[scale=0.35]{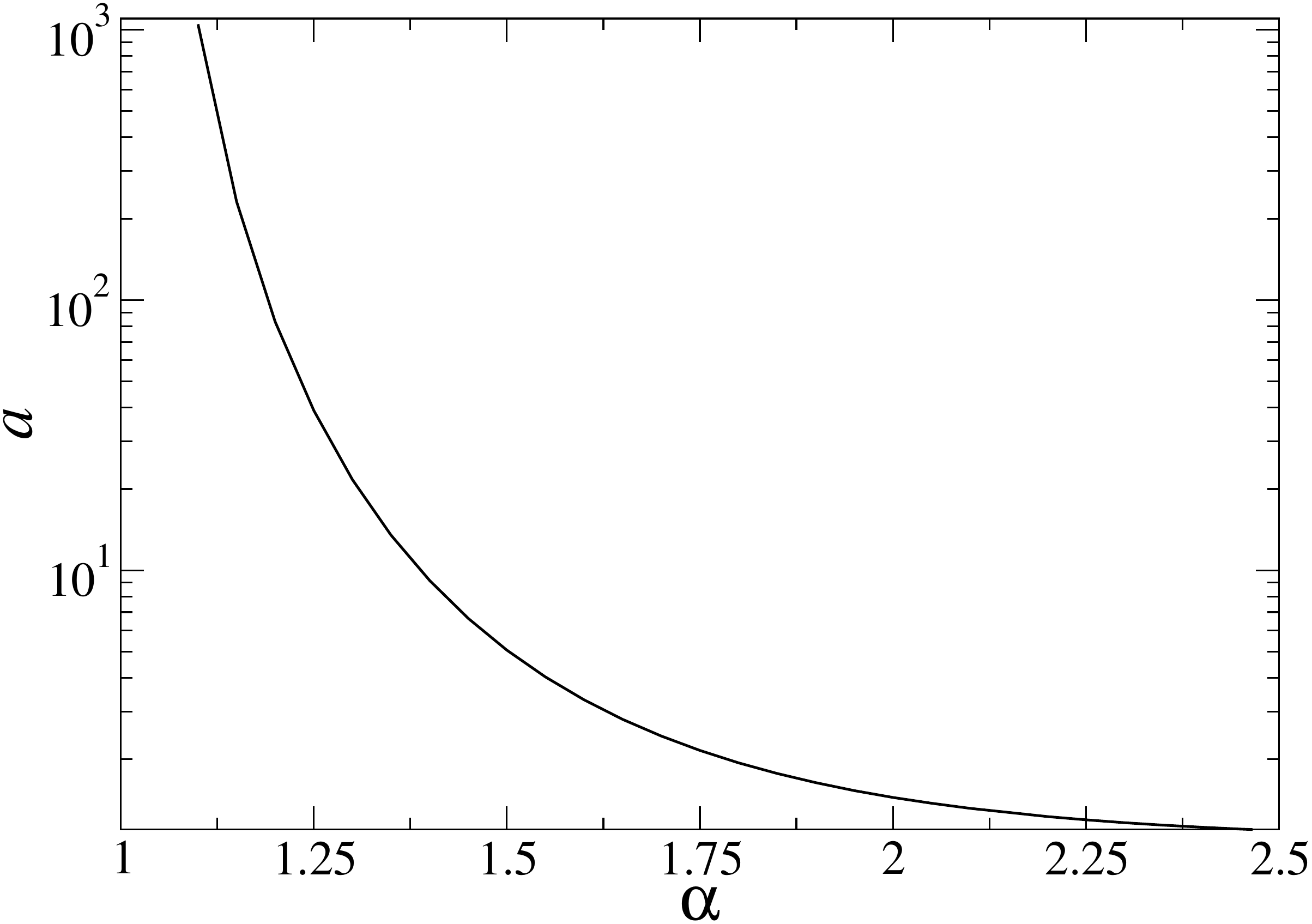}
\caption{(Color online) Plot of $a$ as a function of $\alpha$ obtained by numerically solving  Eq. (\ref{a}) for a
given value of $\beta$ (see Eq. (\ref{SCtrans1})) corresponding to  $D_1=1.3$ and $D_2=1.5$.}\label{fig5bis}
\end{figure}
Using the above expressions in Eq. (\ref{B_0}) we see that $a$ diverges as $a \propto 1/(\alpha - 1)$ as $\alpha$~approaches~$1$. 
Next we consider the case $\alpha \to \infty$ where we expect $a$ to approach $1$. Expanding the functions $h_{\theta}(a)$ and $k_{\theta}(a)$ 
around 
$a =1$ we find
\begin{eqnarray}
 h_{\theta}(a) &\simeq& \pi+\pi(1-\theta)~\tan(\beta)(a-1) + \mathcal{O}\left((a-1)^2\right), \label{hat1} \\
 k_{\theta}(a) &\simeq& -\log(a-1) + \mathcal{O}\left((a-1)\log(a-1)\right),\label{kat1}
\end{eqnarray}
which with the help of Eq. (\ref{B_0}) yields $a -1 \sim e^{-\pi \alpha \tan(\beta)}$. 
In Table \ref{table2} we summarize the values of $a$ and $B_0$ for different $\alpha=L_2/L_1$.
\begin{table}[h]
\centering
\begin{tabular}{|c|c|c|}
\hline
& \; &  \; \\
$\alpha \to 1$ & $a \propto 1/{(\alpha-1)}$ 
& $B_0 = \dfrac{L_1\Gamma[\frac{1}{2}+\theta]\tan{\beta}}{\sqrt{\pi D_1}\Gamma[\theta]}\sqrt{a}e^{i\beta}$ \\ 
& \; &  \; \\
\hline
& \; &  \; \\
$1<\alpha <\infty$ & $1<a <\infty$ & $B_0=\dfrac{L_1\alpha\tan(\beta))}{\sqrt{D_1}k_{\theta}(a)}~e^{i\beta}$ \\
& \; &  \; \\
\hline
& \; &  \; \\
$\alpha \to \infty$ & $a -1 \simeq e^{-\pi \alpha \tan(\beta)}$ & $B_0 = \dfrac{L_1}{\pi\sqrt{D_1}}e^{i\beta}$\\
& \; &  \; \\
\hline
\end{tabular}
\caption{Table of values of $a$ and $B_0$ for given $L_1$, $\alpha={L_2}/{L_1}$ and $\beta=\arctan\left( \sqrt{\frac{D_1}{D_2}}\right)$.}
\label{table2}
\end{table}
Once the values of $B_0$ and $a$ are determined, the conformal transformation in Eq. (\ref{SCtrans2}) is uniquely defined. 
Under this transformation (\ref{SCtrans2}) the Laplace's equation (\ref{laplace1}) remains invariant \emph{i.e} we still have 
\begin{equation}
 \frac{\partial^2 F(x,y)}{\partial x^2}+\frac{\partial^2 F(x,y)}{\partial y^2}=0 \;,
\end{equation}
in the new variables $(x,y)$ which holds over the upper half complex plane. The BCs on 
the real axis are  
\begin{eqnarray}
 ~&&(i)~F(x,0)=0~\text{for}~x < -1~~\text{and}~F(x,0)=1~\text{for}~x \geq-1~~~~~\text{(process 1)} \label{BCP1} \\
 \text{and}~&&(ii)~F(x,0)=1~\text{for}~-1\leq x \leq 0~~\text{and}~F(x,0)=0~ \text{otherwise}~~~~\text{(process 2)} \;. \label{BCP2}
\end{eqnarray}
The solution of the Laplace's equation in the upper half complex plane can be written explicitly in terms 
of the values at the boundary by using Poisson's integral formula
\begin{equation}
 F(x,y)=\frac{y}{\pi}\int_{-\infty}^{\infty} \frac{F(x',0)}{y^2+(x-x')^2} dx' \;.
\end{equation}
Using the BCs in Eqs. (\ref{BCP1}), (\ref{BCP2}) and performing the integral in both cases we get the following explicit solutions 
\begin{eqnarray}\label{solP1}
 F(x,y)= 
 \begin{cases}
 &1 - \frac{1}{\pi}\arctan \left ( \frac{y}{1+x} \right)~~~~~~~~~~~~~~~~~~~~~~\text{for process 1}  \\
&\frac{1}{\pi}\left[ \arctan \left ( \frac{y}{x} \right) -\arctan \left ( \frac{y}{1+x} \right)\right]~~~~~~~\text{for process 2},
 \end{cases}
\end{eqnarray}
expressed in terms of $x$ and $y$. 

\subsection{Results and discussions}
\label{results}
\noindent
In the previous section \ref{confmap}, we have solved the Laplace's equation in~(\ref{laplace1}) using S-C conformal mapping 
which provides the first exit probability $F$ in terms of the $(x,y)$ coordinates 
(\emph{i.e.} in the $z$-plane). Then, to obtain the cumulative distribution $\mathcal{Q}(L_1,L_2|u_1,u_2)$ defined in Eq. (\ref{jntQ}), we first need to express the 
solution $F(x,y)$ in terms of our original coordinates 
$\left(v_1=\frac{u_1}{\sqrt{D_1}},v_2=\frac{u_2}{\sqrt{D_1}}\right)$ which can, in principle, be done by inverting 
the conformal transformation $W(z)=W$ in Eq. (\ref{SCtrans2}). Once this inversion is performed, the marginal cumulative distribution 
$\mathcal{Q}_1(L|u_1,u_2)=\text{Prob.}[m_1\leq L|~0<u_1<L_1;~u_1<u_2<L_{2}]$ is obtained 
by taking $L_1 \to L$ and $L_2 \to \infty$ limits of $\mathcal{Q}(L_1,L_2|u_1,u_2)$ whereas the marginal cumulative distribution 
$\mathcal{Q}_2(L|u_1,u_2)=\text{Prob.}[m_2\leq L|~0<u_1<L_1;~u_1<u_2<L_{2}]$ is obtained 
by taking $L_1 \to L_2=L$ limit of $\mathcal{Q}(L_1,L_2|u_1,u_2)$. 
The inversion of the transformation $W(z)=W$ for any given $L_1$ and $L_2$ can not be done analytically in terms of 
elementary functions but one can do this inversion numerically. 
In the asymptotic limit $L_2 \to \infty$ and $L_1 \to \infty$, as we will show below, the S-C transformation gets simplified and there 
one can invert the transformation analytically. 
In section \ref{numerics} we evaluate $\mathcal{Q}_1(L|u_1,u_2)$ and $\mathcal{Q}_2(L|u_1,u_2)$ obtained by numerical 
inversion of the conformal transformation. These expressions can then be compared to their numerical estimation obtained by direct 
simulation of Langevin equations (\ref{lang-2-part}). In section \ref{asymptotes} we present large $L$ asymptotics which allows to study the 
tails of the marginal distributions $p_i(m|u_1,u_2)$. Finally, in section \ref{correl} we study the correlations between $m_1$ 
and $m_2$. 
\subsubsection{Evaluation of $\mathcal{Q}_1(L|u_1,u_2)$ and $\mathcal{Q}_2(L|u_1,u_2)$ through numerical inversion}
\label{numerics}
\begin{figure}[ht]
\centering
\includegraphics[scale=0.5]{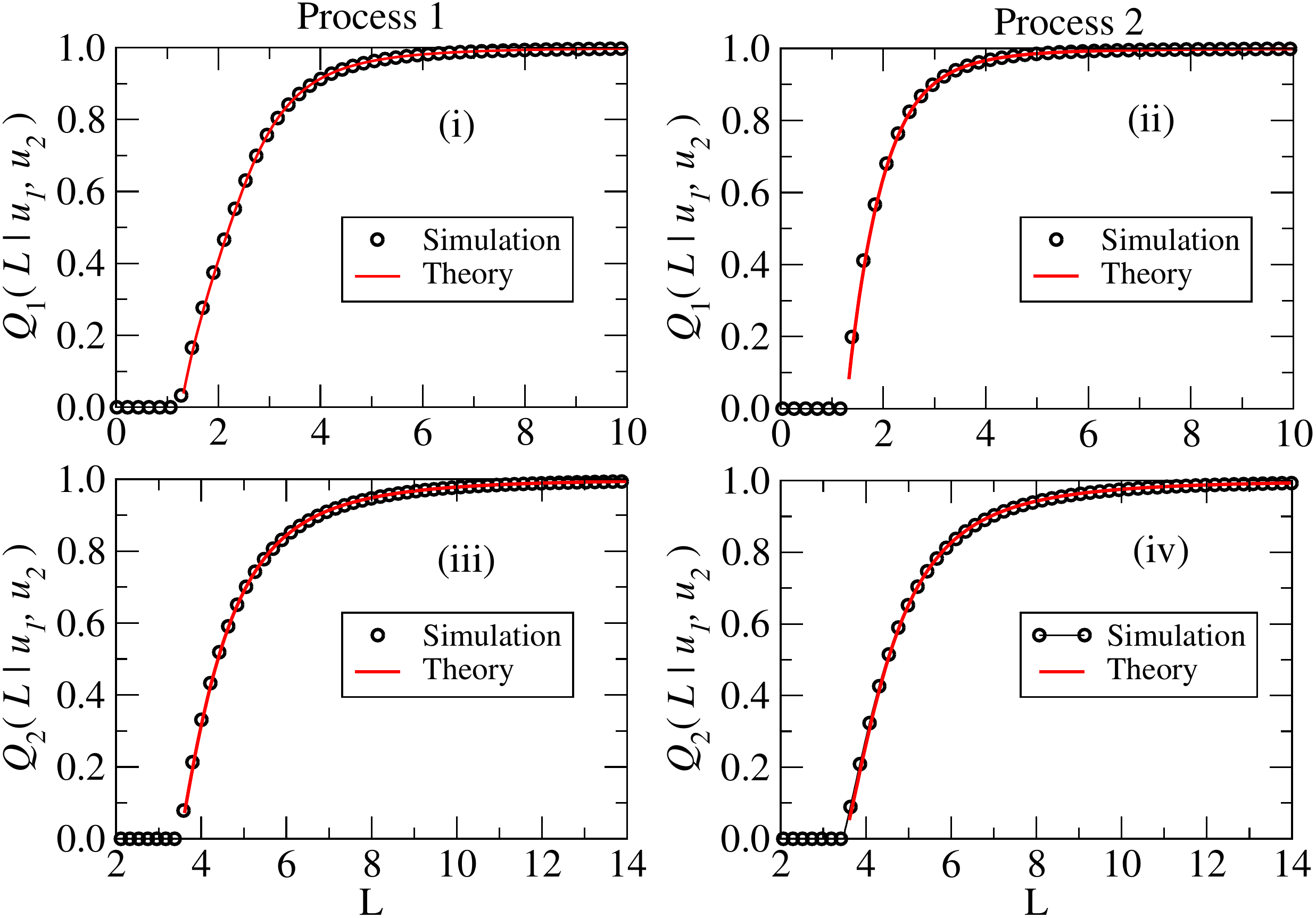}
\caption{ (Color online) Plots of marginal cumulative distributions. The open circles represent the data obtained from numerical simulations while 
the solid line corresponds to a numerical evaluation of our exact formula. In figures (i) and (ii) we show a plot of 
$\mathcal{Q}_1(L|u_1,u_2)$ as a function of $L$ while in figures (iii) and (iv) we show a plot of $\mathcal{Q}_2(L|u_1,u_2)$ again as a function  $L$ 
for both processes. The parameters used for this plot are 
$u_1=1.27,~u_2=3.51$,$D_1=1.3$ and $D_2=1.5$.}
\label{fig6}
\end{figure}

 We first compute $\mathcal{Q}_1(L|u_1,u_2)$, \emph{i.e.} the probability that the maximum $m_1$ of the 1st particle remains below
 $L$ till the stopping of the two-particle process. It is obtained from $\mathcal{Q}(L_1,L_2|u_1,u_2)$ in the $L_2 \to \infty$ limit 
 keeping $L_1=L$ fixed. This corresponds to $\alpha = L_2/L_1 \to \infty$ which, 
with the help  of table \ref{table2}, implies $a \to 1$ and $B_0=\frac{L}{\pi\sqrt{D_1}}e^{i\beta}$ where $\beta$ is given in Eq. (\ref{SCtrans1}). 
Using this value of $B_0$ and taking the limit $a \to 1$ of the S-C transformation in Eq. (\ref{SCtrans2}) 
we have 
\begin{equation}
 \frac{\pi\sqrt{D_2u_1^2+D_1u_2^2}}{L~\sqrt{D_2}}~e^{ i\left(\psi-\beta \right)}
 = \int_0^z \frac{t^{\theta-1}}{(1+t)}~dt, \label{SCtrans3}
\end{equation}
where we have written the complex coordinate $W=\left(\frac{u_1}{\sqrt{D_1}} + i \frac{u_2}{\sqrt{D_2}} \right)$ on the left hand side as 
\begin{equation}
 W= \frac{\sqrt{D_2u_1^2+D_1u_2^2}}{\sqrt{D_1D_2}}e^{i\psi}, ~~\text{with}~~
 \psi=\arctan\left( \frac{\sqrt{D_1}u_2}{\sqrt{D_2}u_1}\right).
\end{equation}
For given $(u_1, u_2)$ and $L$, we numerically solve the above equation (\ref{SCtrans3})
for $z=x+iy$. Plugging this solution into Eq. (\ref{solP1}) first and then using Eq. (\ref{P0}) we obtain $\mathcal{Q}_1(L|u_1,u_2)$. 
In Fig. \ref{fig6} (i) and (ii) we compare the value of $\mathcal{Q}_1(L|u_1,u_2)$ obtained from numerical inversion of the transformation $W(z)$ 
[i.e. solving Eq. (\ref{SCtrans3})] to the value of $\mathcal{Q}_1(L|u_1,u_2)$ obtained from direct simulation of the Langevin Eqs.~(\ref{lang-2-part}) 
for both process 1 and process 2, and for $u_1=1.27,~u_2=3.51$, $D_1=1.3$ and $D_2=1.5$. We observe a very good agreement between the analytical and 
numerical curves.   

Similarly, to evaluate $\mathcal{Q}_2(L|u_1,u_2)$, \emph{i.e.} the probability that the maximum $m_2$ of the 2nd particle stays below $L$ till $t_s$, 
we first take the limit $L_2 \to L_1=L$ (\emph{i.e.} $\alpha \to 1$) of $\mathcal{Q}(L_1,L_2|u_1,u_2)$. We know from Table \ref{table2} that, 
when $\alpha \to 1$ the coordinate $a$ goes to $\infty$. As a result, the S-C transformation in Eq. (\ref{SCtrans2}) now reads 
\begin{equation}\label{invertW_2}
 \frac{\Gamma[\theta]\Gamma[\frac{1}{2}-\theta]\sqrt{D_2u_1^2+D_1u_2^2}}{L~\sqrt{\pi}\sqrt{D_1+D_2}}~e^{ i\left(\psi-\beta \right)}
 = \int_0^z \frac{t^{\theta-1}}{\sqrt{1+t}}~dt \;. 
\end{equation}
For given $(u_1,~u_2)$ and $L$, we solve the above equation 
for $z=x+iy$ numerically and plug the solution into Eq. (\ref{solP1}) to obtain $\mathcal{Q}_2(L|u_1,u_2)$ from Eq. (\ref{P0}).
In Fig. \ref{fig6} (iii) and (iv) we 
compare the value of 
$\mathcal{Q}_2(L|u_1,u_2)$, obtained by numerically inverting the transformation $W(z)$ [i.e. solving Eq. (\ref{invertW_2})] to the value of 
$\mathcal{Q}_2(L|u_1,u_2)$ obtained from direct simulation of the Langevin Eqs. (\ref{lang-2-part}) for both 
process 1 and process 2, and for $u_1=1.27,~u_2=3.51$, $D_1=1.3$ and $D_2=1.5$: here also one observes a very good agreement between the 
analytical and numerical curves.

\subsubsection{Large $L_1$ and $L_2$ limits }
\label{asymptotes}
\noindent
We now focus on the limit where both $L_1$ and $L_2$ are large, keeping the ratio $\alpha = L_2/L_1$ fixed. In this limit, 
the S-C transformation gets simplified which makes it possible to invert the transformation $W=W(z)$ analytically.
For simplicity we present here our calculation assuming $D_1=D_2=D$. The calculation for $D_1 \neq D_2$ can be done similarly. For $D_1=D_2=D$ we have 
\begin{equation}
\beta=\frac{\pi}{4},~~\theta=\frac{1}{4},~~W=\frac{u_1+i~u_2}{\sqrt{D}}=\frac{\sqrt{u_1^2+u_2^2}}{\sqrt{D}}e^{i\psi}~~
\text{where}~~\psi=\arctan\left(\frac{u_2}{u_1}\right) \;.
\label{psi}
\end{equation}
From Eq. (\ref{B_0}), we see that $B_0$ diverges linearly with $L_1$ as $k_{\theta}(a)$ is finite (see Fig. \ref{fig5} ii). 
Dividing both sides of Eq.~(\ref{SCtrans2}) by $B_0$ and 
taking $L_1 \to \infty$ and $L_2 \to \infty$ limit while keeping $\alpha=L_2/L_1$ fixed, we see the left hand side of Eq. (\ref{SCtrans2}) decreases to zero. 
This suggests us to expand the integral on the right hand side 
of Eq. (\ref{SCtrans2}) around $z=0$ to get 
\begin{equation}
\frac{W}{B_0} =\frac{\left(u_1 + i u_2 \right)}{B_0\sqrt{D}} 
= \frac{z^{\frac{1}{4}}}{\sqrt{a}}\left [4 - \frac{2(1+a)}{5a}z + \mathcal{O}(z^2)\right]. 
\end{equation}
Following Ref. \cite{SatyaBray2010}, we now invert the transformation $W=W(z)$ from $z$-plane to $(u_1,u_2)$ plane 
to obtain $z \approx \mathcal{R}e^{i\Psi}$ where, denoting $L_1=L$ and $L_2=\alpha L$, we have
\begin{eqnarray}
 \mathcal{R} &=& \left[\mathcal{X}(a)\frac{(u_1^2+u_2^2)^2}{L^{4}} 
 +\frac{2(1+a)\mathcal{X}(a)^2}{5a} \frac{(u_1^2+u_2^2)^4\cos\left(4\psi-\pi\right)}{L^{8}}
  + \mathcal{O}(L^{-12}) \right],
 \label{z} \allowdisplaybreaks[4] \\
\Psi&=&4\psi-\pi+\frac{2(1+a)\mathcal{X}(a)}{5a}\frac{(u_1^2+u_2^2)^2\sin\left(4\psi-\pi\right)}{L^{4}}
+ \mathcal{O}(L^{-12}) , \label{Psi} \allowdisplaybreaks[4] \label{Cpsi}\\
&&\text{with} \; \mathcal{X}(a)=\left (\frac{h_{1/4}(a)\sqrt{a}}{4} \right)^{4}, 
\label{mcalX}
\end{eqnarray}
and $h_{1/4}(a)$ is defined in Eq.~(\ref{hka}). This large $L$ expansion can in principle be carried out systematically to arbitrary order. We now take the  
small $z$ limit of the explicit solutions $F(x,y)$ in 
Eq. (\ref{solP1}) and then inject the above large $L$ expansion of $z \approx \mathcal{R}e^{i\Psi}$ into it 
to get the exit probability, mentioned in Eq. (\ref{P0}), as 
\begin{eqnarray}
\label{SAP1}
 F(u_1,u_2;L,\alpha L) \approx
 \begin{cases}
&1-\dfrac{\mathcal{X}(a)}{\pi}(u_1^2+u_2^2)^2\sin\left(4\psi-\pi\right)L^{-4}
 +\mathcal{O}(L^{-8}) \;\;\;\;\hspace*{0.35cm} \text{for process 1} \;,  \\
 &\\
 & \dfrac{4\psi-\pi}{\pi}-\dfrac{1}{\pi}\left(\dfrac{3a-2}{5a}\right)\mathcal{X}(a)
 (u_1^2+u_2^2)^2\sin\left(4\psi-\pi\right)L^{-4}+ \mathcal{O}(L^{-12}) \\
 &\hspace*{8cm}\text{for process 2},
 \end{cases}
\end{eqnarray}
where $\psi$ and $\mathcal{X}(a)$ are given in Eq. (\ref{psi}) and (\ref{mcalX}) respectively, with $a$ implicitly determined from Eq. (\ref{a}). 
Putting $L=\infty$ in the above equation (\ref{SAP1}) we get the probability $S_2(u_1,u_2)$ defined in Eq. (\ref{Ult-Survivl}), which for process 1 is equal to $1$ 
and for process 2 is equal to $\frac{4\psi-\pi}{\pi}$. Using the expression of $\psi$ from Eq. (\ref{psi}), we get explicit expression of 
$S_2(u_1,u_2)$ for process 2, as announced in Eq. (\ref{Q}) with $D_1=D_2$. 
One can follow the same calculation to get $S_2(u_1,u_2)$ for $D_1 \neq D_2$. After few simplifications, one can 
rewrite the exit probability $F(u_1,u_2;L,\alpha L)$ in Eq. (\ref{SAP1}) in the following form :
\begin{equation}
F(u_1,u_2;L,\alpha L) \approx S_2(u_1,u_2) - c(a) \mathcal{Y}_2(u_1,u_2)~L^{-4}~~\text{for}~L\gg u_2, \label{F} 
\end{equation}
for both processes 1 and 2, where the constant $c(a)$ is given by 
\begin{eqnarray}\label{c(a)}
 c(a) = 
 \begin{cases}
  & \dfrac{4\mathcal{X}(a)}{\pi},~~~~~~~~~~~~~~~~\text{for~process 1}\\
  & \\
  &\dfrac{4\mathcal{X}(a)}{\pi}\left(\dfrac{3a-2}{5a}\right),~~\text{for~process 2},
 \end{cases}
\end{eqnarray}
\begin{equation}
 \text{and}~~\mathcal{Y}_2(u_1,u_2) = u_1u_2(u_2^2-u_1^2)~\text{for~both~processes~1~and~2.}\label{Vfunc}
\end{equation}
Plugging the expression of $F(u_1,u_2;L,\alpha L)$ from Eq.~(\ref{F}) into 
Eq.~(\ref{P0}), we get the joint cumulative distribution 
\begin{eqnarray}
\mathcal{Q}(L,\alpha L|u_1,u_2) &\simeq & 1- c(a) \frac{\mathcal{Y}_2(u_1,u_2)}{S_2(u_1,u_2)}~\frac{1}{L^4},~~\text{for}~L \gg u_2~\;. \label{Q-i} 
 \end{eqnarray}
Note that the $\alpha$ dependence in the above expression comes only through $a$ since it is a function of $\alpha=\frac{L_2}{L_1}$ 
[see Eq.~(\ref{a})]. 
Taking the limit $\alpha \to \infty$ \emph{i.e.} $a \to 1$ (see Table \ref{table2}) in the above expression, we get the
marginal cumulative distribution 
$\mathcal{Q}_1(L|u_1,u_2)$ [defined below Eq. (\ref{jntQ})] of the maximum $m_1$ given that the 1st particle always stayed below the 2nd
particle till the stopping time $t_s$. Similarly, if we take the limit $\alpha \to 1$ \emph{i.e.} $a \to \infty$ limit, we get 
$\mathcal{Q}_2(L|u_1,u_2)$, the cumulative distribution of the maximum $m_2$. 
After taking the derivative 
of $\mathcal{Q}_i(L|u_1,u_2)$ with respect to $L$ and putting $L=m$ we get the marginal PDFs, $p_i(m|u_1,u_2)$ for 
$i=1,2$ which behave like
\begin{eqnarray}
&& p_1(m|u_1,u_2) \approx k_1 {\mathcal{Y}_2(u_1,u_2)}\frac{1}{m^5},~\text{for}~m\gg u_2 \;, \\
&& p_2(m|u_1,u_2) \approx k_2\frac{\mathcal{Y}_2(u_1,u_2)}{S_2(u_1,u_2)}\frac{1}{m^5},~\text{for}~m\gg u_2 \;,\label{p2-tail}
\end{eqnarray}
where the numerical constants $k_1$ and $k_2$ are obtained by taking, respectively, $a\to 1$ and $a \to \infty$ limits of $c(a)$ and finally 
multiplying it by $4$ [coming from the derivative of $L^{-4}$ w. r. t. $L$ in (\ref{F})]. The constants $k_1$ and $k_2$ are explicitly given by  
\begin{eqnarray}
\label{k-consts}
k_1 = 
\begin{cases}
&\dfrac{\pi^3}{16} \;, \; \text{for~process 1} \\
& \\
&\dfrac{\pi^3}{80} \;, \; \text{for~process 2}
\end{cases}\hfill \hspace*{1cm} {\rm and} \; \hspace*{1cm}
k_2 = \begin{cases}
& \dfrac{\Gamma[\frac{1}{4}]^8}{(4\pi)^3}\;,~~~~\text{for~process 1} \\
& \\
&  \dfrac{3~\Gamma[\frac{1}{4}]^8}{5~(4\pi)^3}\;,~~\text{for~process 2} \;.
\end{cases}
\end{eqnarray}
We can easily see that the tails of the PDFs $p_i(m|u_1,u_2)$ in Eq.~(\ref{p2-tail}) 
are of the form announced in Eqs.~(\ref{2part-dist-intro1}) and (\ref{2part-dist-intro2}) with $D_1=D_2$. 
The Vandermonde determinant in the expression of $\mathcal{Y}_2(u_1,u_2)$~(\ref{Vfunc}) reflects the fact that the two walkers are non-intersecting and is reminiscent of the connection between vicious walkers and 
random matrix theory \cite{Schehr08}. 

A similar calculation can be performed in the case of different diffusion constants $D_1 \neq D_2$ to obtain
\begin{equation}
  \mathcal{Q}(L,\alpha L|u_1,u_2) \approx 1 - \frac{\mathcal{B}(u_1,u_2,\alpha, \mu)}{L^{\mu}} ~~\text{with}~~
  \mu=\frac{2\pi}{\pi-2\arctan \left( \sqrt{\frac{D_1}{D_2}} \right)}~~~\text{for}~~L\gg u_2\label{unequalD}
\end{equation} 
which finally provides the PDFs 
\begin{equation}
p_i(m|u_1,u_2) \approx \frac{\mathcal{A}_i(u_1,u_2,\mu)}{m^{\nu}} ~~\text{with}~~\nu=\mu+1~~~\text{for}~~m\gg u_2 \;. \label{punequalD}
\end{equation}
The explicit expressions of the functions $\mathcal{B}(u_1,u_2,\alpha, \mu)$ and $\mathcal{A}_i(u_1,u_2, \mu)$ have been left in 
Appendix \ref{expreuneqD}. Here we present the results of $p_1(m_1|u_1,u_2)$ and $p_2(m_1|u_1,u_2)$ obtained by simulating directly the Langevin 
equations in Eq. (\ref{lang}) and compare them with the analytical prediction in Eq. (\ref{punequalD}).
\begin{figure}[ht]
\centering
\includegraphics[scale=0.5]{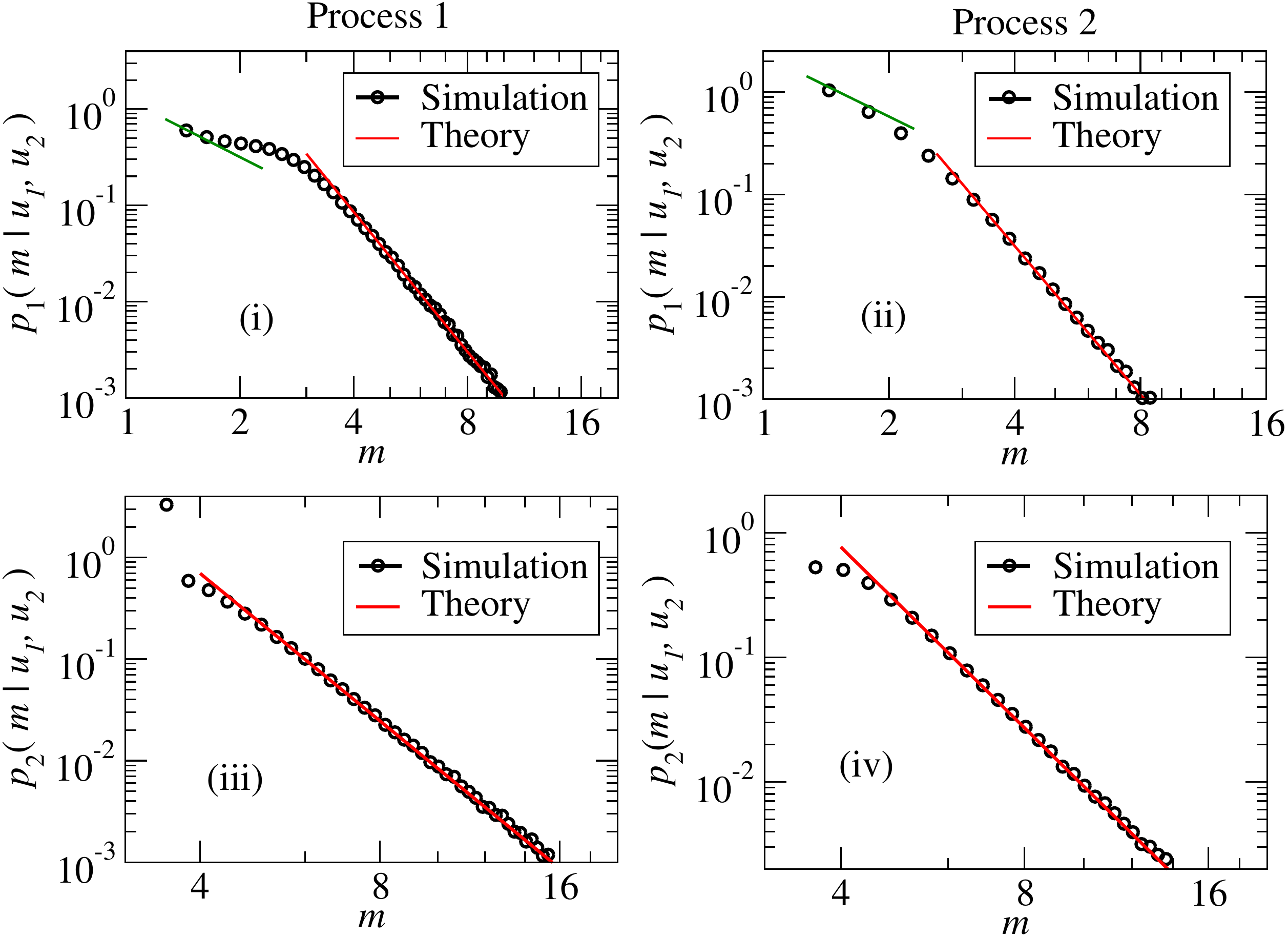}
\caption{(Color online) Plots of the PDFs of $m_1$, the maximum of the first (left) particle and of $m_2$, the maximum of the 
second (right) particle up to the stopping time $t_s$. The open circles correspond to numerical simulations while the solid lines represent the 
exact asymptotic behaviors (the red ones for the large argument asymptotics and the green one for the small argument asymptotics) as 
explained in the text. There is no fitting parameters. In figures (i) and (ii) we show a plot of 
$p_1(m|u_1,u_2)$ as a function of $m$ whereas, in figures (iii) and (iv) we show a plot of $p_2(m|u_1,u_2)$ as a function of $m$ 
for both processes. The parameters used for these plots are $u_1=1.27,~u_2=3.51$, $D_1=1.3$ and $D_2=1.5$. The red solid lines have a 
slope $-\nu = -4.826$ as expected from Eqs. (\ref{unequalD}) and (\ref{punequalD}) while the green solid lines have a slope $-2$.}
\label{fig7}
\end{figure}
In Fig. \ref{fig7} (i) and (ii) we show a plot $p_1(m|u_1,u_2)$ (with open circles) obtained from simulation, respectively for process 1
and process 2, with $u_1=1.27,~u_2=3.51$, $D_1=1.3$ and $D_2=1.5$. We have also plotted the large $m$ asymptotic behavior
obtained from our analytical prediction (\ref{punequalD}) with the same set of parameters for which, 
one expects from Eqs. (\ref{unequalD}) and 
(\ref{punequalD}) that $\nu \simeq 4.826$ and, from Eqs. (\ref{AaP11}) and (\ref{AaP12}), $\mathcal{A}_1(u_1,u_2,\mu) \simeq 68.79$ for process 1 
and 
$\mathcal{A}_1(u_1,u_2,\mu)  \simeq 25.32$ for process 2. We see that the agreement between our numerical simulations and our analytical results is 
very good. Notice also that for $m_1 \ll u_2$, the first particle does not feel the presence of the second particle and therefore one expects 
that in this limit  $p_1(m|u_1,u_2) \sim {u_1}/{m^2}$ for process 1 and $p_1(m|u_1,u_2) \sim {u_1}/{[S_2(u_1,u_2)m^2]}$ for process 2, with 
$S_2(u_1,u_2) \simeq 0.549$ 
in the later case [see Eq.~(\ref{Q})]. These asymptotic behaviors for small $m$ are shown as green line in Fig. \ref{fig7} (i) and (ii): the agreement 
with the numerical data is rather good. Finally, in Fig. \ref{fig7} (iii) and (iv) we show a comparison between $p_2(m|u_1,u_2)$ evaluated 
numerically (open circles) and the analytical predictions for the tails (\ref{punequalD}) [see also Eqs. (\ref{AaP21}) and (\ref{AaP22})]. Here also 
the agreement is very good.

\subsubsection{Correlation between the maxima $m_1$ and $m_2$}\label{correl}

We end up this section by considering the correlations between the maximal displacements $m_1$ and $m_2$ of the 1st and 2nd particle, respectively.
To characterize these correlations, we define the following quantity 
\begin{equation}
 C_{\theta}(\alpha,L ; u_1,u_2)=[\mathcal{Q}(L,\alpha L| u_1,u_2)-\mathcal{Q}_1(L|u_1,u_2)\mathcal{Q}_2(\alpha L|u_1,u_2)]L^{\frac{1}{\theta}}, \label{Cmea}
\end{equation}
with $\alpha = L_2/L_1$ and $\theta$ is given in Eq. (\ref{SCtrans2}). This quantity 
measures the difference between the joint cumulative probability $\mathcal{Q}(L,\alpha L| u_1,u_2)$ of $m_1,m_2$ and the product of their individual 
marginal cumulative probabilities $\mathcal{Q}_1(L|u_1,u_2)$ and $\mathcal{Q}_2(\alpha L|u_1,u_2)$. 
\begin{figure}[t]
\centering
\includegraphics[scale=0.4]{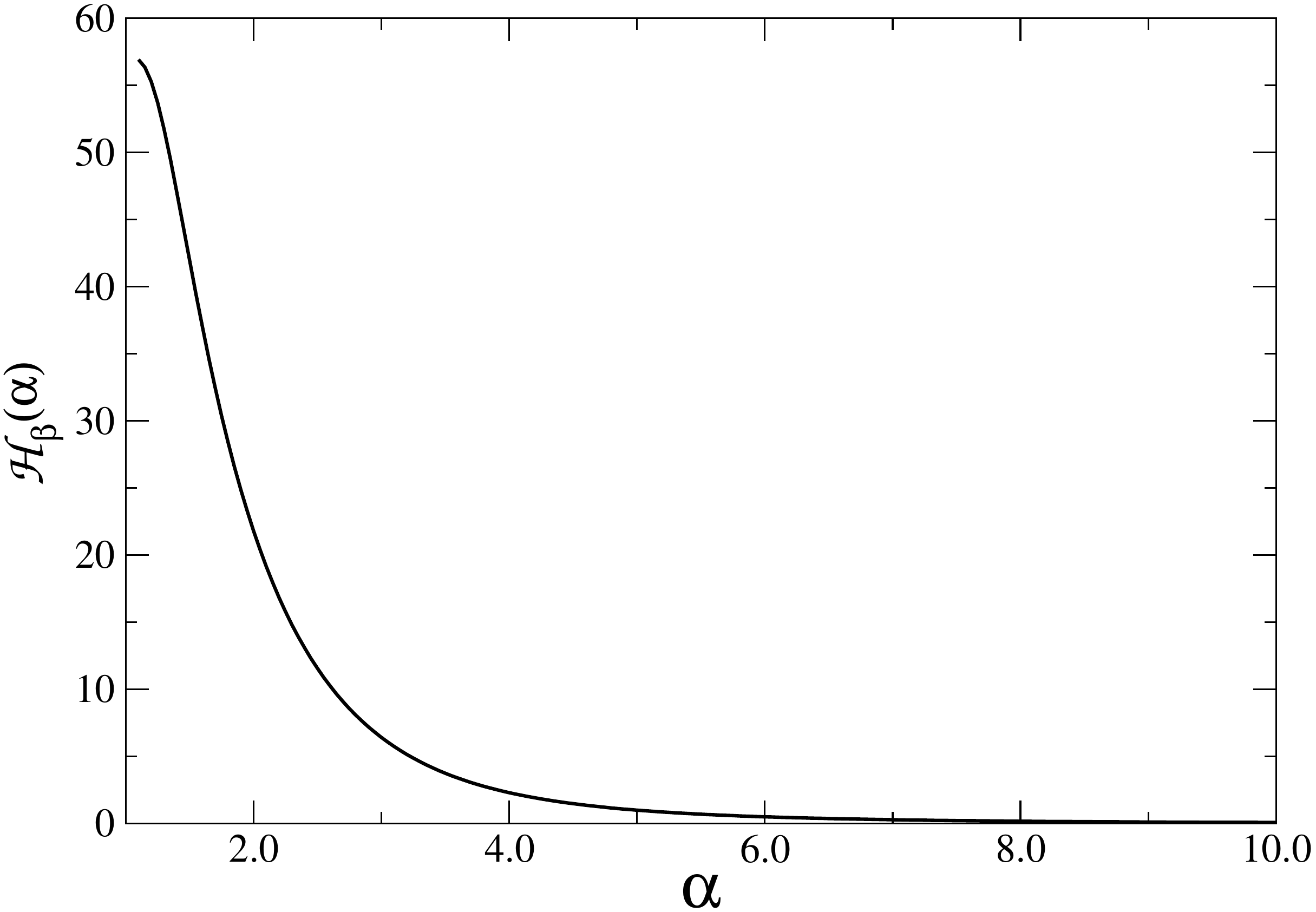}
\caption{(Color online) Plot of $\mathcal{H}_{\beta}(\alpha)$, which gives a measure of correlation of the two maxima $m_1$ and $m_2$ for 
$D_1=1.3$ and $D_2=1.5$.}
\label{fig8}
\end{figure}
If the two maxima $m_1$ and $m_2$ are independent of each other then the quantity defined above would be identically zero for any 
$\alpha$ and $L$. 
We plug the large $L$ expression of $\mathcal{Q}(L,\alpha L| u_1,u_2)$ from Eq. (\ref{unequalD}) and the large $L$ of 
$\mathcal{Q}_1(L|u_1,u_2)$ and $\mathcal{Q}_2(\alpha L|u_1,u_2)$ 
obtained from Eq. (\ref{unequalD}) by taking $\alpha \to \infty$ and $\alpha \to 1$ limit, respectively, to see that 
the function $C_{\theta}(\alpha,L ; u_1,u_2)$ becomes independent of $L$ [the factor $L^{1/\theta}$ 
in (\ref{Cmea}) is chosen for this purpose] and takes the following form
\begin{equation}
 C_{\theta}(\alpha,L ; u_1,u_2)|_{L\to \infty}= \mathcal{H}_{\beta}(\alpha) \mathcal{L}_{\beta}(u_1,u_2) \;,
\end{equation}
for both processes, where the function $\mathcal{H}_{\beta}(\alpha)$ carries the information on the correlations 
(the function ${\cal L}_\beta(u_1,u_2)$ can also be computed explicitly but we do not discuss it here). 
This function $\mathcal{H}_{\beta}(\alpha)$ is given by
\begin{eqnarray}
 \mathcal{H}_{\beta}(\alpha)&=& 
 \frac{1}{\pi}\left[\left (\pi \tan(\beta)\right )^{\frac{1}{\theta}} + \alpha^{-\frac{1}{\theta}}
 \left(\frac{\sqrt{\pi}\Gamma[\theta]}{\Gamma[\frac{1}{2}+\theta]} \right)^{\frac{1}{\theta}} 
 - \left(\frac{k_{\theta}(a)\sqrt{a}}{\alpha} \right)^{\frac{1}{\theta}}\right], \nonumber \\
 \text{with}&& \theta = \frac{1}{2} - \frac{\beta}{\pi}~~\text{and}~~\beta=\arctan\left( \frac{\sqrt{D_1}}{\sqrt{D_2}}\right).
\end{eqnarray}
Here we should keep in mind that according to Eq. (\ref{a}), $a$ is a function of $\alpha$ and $\beta$. 
In Fig. \ref{fig8}, we plot $\mathcal{H}_{\beta}(\alpha)$ as a function of $\alpha$ for $D_1=1.3$ and $D_2=1.5$. This shows that even when both 
$m_1,m_2$ are large, $m_1$ and $m_2$ are strongly correlated as long as they are of the same order of magnitude. As expected, 
these correlations vanish when the two particles are very far away from each other.


\section{Multi-particle problem: $N>2$}
\label{multipart}
In this section we generalize the two vicious walkers problem to $N$ vicious walkers problem. 
We focus on the PDF $p_N(m|{\bf u})$ of the global maximum $m_N$ 
(maximal distance travelled by the rightmost particle). Moreover, we assume that the $N$ walkers are identical \emph{i..e.} 
they have the same diffusion constant $D_1=D_2=...=D_N=D$. In this case, we expect that the PDF $p_N(m|{\bf u})$ will not depend on $D$ 
and  will have the following power law tail 
\begin{equation}
 p_N(m|{\bf u}) \simeq \dfrac{\mathcal{A}_N({\bf u})}{m^{\nu_N}}~~\text{with}~\nu_N=N^2+1, \label{P_N_multi}
\end{equation}
for both processes 1 and 2.
We first show from a heuristic scaling argument that one can predict 
the power law in the above equation. 
The scaling argument is based on the large time tail of the distribution $f_N(t_s|{\bf u})$ of the stopping time $t_s$. 
This argument provides the $N$-dependence of the exponent $\nu_N$ accurately but it does not predict the prefactor precisely. For this we study the 
$N$-particle problem rigorously in the next subsection, where we follow an approach different from what we have done for the $N=2$ case. 
In particular, we have used the Green's function 
approach directly rather than solving a $N$-dimensional Laplace's equation inside an $N$-dimensional complicated Weyl chamber because the later approach becomes difficult as 
we do not have at our disposal any generalized Schwarz-Christoffel transformation valid in dimensions $d>2$.

\subsection{A heuristic argument for $N$ particles}
\label{argument}
\noindent 
To justify the power law for $p_N(m|{\bf u})$ given in Eq.~(\ref{P_N_multi}), we present a simple scaling argument which is based 
on the power law tail of the PDF of the 
stopping time $t_s$ itself. This argument is valid for both process 1 and process 2 and it goes as follows. 
Let $Q_N(L|{\bf{u}})=\text{Prob.}[m_N \leq L|{\bf{u}}]$ be the probability that the 
global maximum $m_N$ stays below the level $L$ given that the $N$ walkers starting from positions ${\bf{u}}$ stay non-intersecting till time $t_s$. For large $L$, 
the dominant contribution to $1-Q_N(L|{\bf{u}}) = \text{Prob.}[m_N \geq L|{\bf{u}}]$ comes from trajectories
which typically have large global maximum $m_N$. On the other hand,
using the connection between non-intersecting Brownian motions and random matrix theory (RMT) one can argue that the average value of the global
maximum $m_N$ over the
time interval $[0,t_s]$ grows as $\langle m_N \rangle \sim \sqrt{t_s~N}$, whereas the  
fluctuations of $m_N$ around this mean value decays as $N^{-1/6}$ \cite{Schehr08,Forrester08}. As a result, the distribution of $m_N$ over the
time interval $[0,t_s]$
will be highly peaked around $m_N \sim \sqrt{t_s N}$ for very large $N$. Therefore one expects that for large $t_s$ and $N$, 
the random variable $m_N$ will be typically of the order of $\sim \sqrt{t_s~N}$. Hence, for large $N$ and $L$, the tail of the 
cumulative distribution 
$Q_N(L|{\bf{u}})$ is obtained from :
\begin{eqnarray}
 1-Q_N(L|{\bf{u}}) &=& \text{Prob.}[m_N > L|{\bf{u}}] \approx \text{Prob.}\left[t_s > c \frac{L^2}{N} \big{|} {\bf{u}}\right], \label{scaling1} \\
 \text{with},&&\text{Prob.}[t_s > t|{\bf{u}}] = \int_t^{\infty} f_N(t'|{\bf{u}}) dt' \;, \label{scaling2}
\end{eqnarray}
where $f_N(t_s|{\bf{u}})$ is the PDF of the stopping time $t_s$ (and $c$ is an undetermined constant, irrelevant for the present argument). 

To find the PDF of the ``stopping time'' $f_N(t|{\bf{u}})$ for identical walkers  
\emph{i.e.} for $D_1=D_2=...=D_N=D$, 
we start with the Green's function $G_N({\bf{y}},t; {\bf{u}},0)$ of  
$N$ non-intersecting Brownian walkers with an absorbing wall at the origin. This Green's function 
represents the probability density of the positions ${\bf{y}}=(y_1,y_2,...,y_N)$ of the $N$ walkers at time 
$t$ given that they had started from positions ${\bf{u}}=(u_1,u_2,...,u_N)$ initially. 
Using the Karlin-McGregor formula \cite{karlin}, this $N$-particle Green's function can be expressed 
as the determinant of a $N\times N$ matrix $[G_{i,j}]\equiv [g(u_i,y_j,t)]$ where 
\begin{equation}
 g(u,y,t)=\frac{1}{\sqrt{4\pi Dt}}\left(\text{exp}\left[-\frac{(y-u)^2}{4Dt}\right] -\text{exp}\left[-\frac{(y+u)^2}{4Dt}\right]\right) \label{gsingle}
\end{equation}
is the single particle Green's function with an absorbing wall at the origin. One can also map the 
problem of finding $G_N({\bf{y}},t; {\bf{u}},0)$ to the problem of finding the wave function of 
$N$ free fermions with an infinite wall at the origin and this mapping allows us to write \cite{Schehr08} 
\begin{equation}
 G_N({\bf{y}},t|{\bf{u}},0) = \frac{1}{(\sqrt{2Dt})^N}\frac{1}{N!}\left( \frac{2}{\pi}\right)^{N}
 \int_0^{\infty}\dots \int_0^{\infty}dq_1dq_2\dots dq_N ~\Phi_{{\bf{q}}}^{(N)}\left(\frac{{\bf{u}}}{\sqrt{2Dt}} \right)
 \Phi_{{\bf{q}}}^{(N)}\left(\frac{{\bf{y}}}{\sqrt{2Dt}}\right) e^{-\frac{1}{2}{\bf{q}}^2} \;, \label{Green1}
\end{equation}
where ${\bf{q}} \equiv \{q_1,q_2,\dots, q_N\}$, ${\bf q}^2=\sum_{i=1}^Nq_i^2$  and 
\begin{equation}
\label{phi}
 \Phi_{{\bf{q}}}^{(N)}\left({\bf{u}} \right) = \left|
 \begin{matrix}
  \sin\left( q_1u_1\right) & \sin\left( q_1u_2\right) & \cdots & \sin\left( q_1u_N\right) \\
  \sin\left( q_2u_1\right) & \sin\left( q_2u_2\right)& \cdots & \sin\left( q_2u_N\right) \\
  \vdots  & \vdots  & \ddots & \vdots  \\
  \sin\left( q_Nu_1\right) & \sin\left( q_Nu_2\right) & \cdots &\sin\left( q_Nu_N\right)
 \end{matrix}\;\right|.
\end{equation} 

In case of process 1, the motion of the $N$ walkers ``stop'' when either any two particles meet each other for the first time before the leftmost 
particle hits the origin or 
the first particle crosses the origin for the first time before any two particles meet each other. 
The survival probability $S_N\left(t|{\bf{u}}\right)$ of such $N$-particle process, is given by 
\begin{equation}
 S_N\left(t|{\bf{u}}\right) = \int_{\mathcal{W}}d^N{\bf{y}}~ G_N({\bf{y}},t|{\bf{u}},0).
\end{equation}
The Weyl chamber $\mathcal{W}$ in the above expression is defined as
$\mathcal{W}=[{\bf y} \in \mathbb{R}_+^N |0\leq y_1\leq y_2 \leq \dots \leq y_N \leq \infty]$ where $\mathbb{R}_{+}$ is the set of non-negative real numbers.
Hence, for process 1, the PDF of the 
``stopping time'' is 
\begin{equation}
f_N(t|{\bf{u}}) =- \frac{\partial S_N\left(t|{\bf{u}}\right)}{\partial t}=  -\frac{\partial }{\partial t} 
 \left( ~\int_{\mathcal{W}} dy_1 dy_2 \dots dy_N~ G_N({\bf{y}},t|{\bf{u}},0)\right)~~~~
 \text{for~process 1}.\label{f-p1}
 \end{equation}
On the other hand, for process 2 the reasoning is a bit different as
the process gets ``stopped'' only when the first particle hits the 
origin for the 
first time before any two other particles collide. This implies that
the PDF $f_N(t|{\bf{x}})$ of  the ``stopping time'' is obtained from the outward flux through the $y_1=0$ hyperplane 
$\widetilde{\mathcal{W}}=[{\bf y} \in \mathcal{W}~|~y_1=0]$ of the Weyl chamber $\mathcal{W}$. Hence integrating the outward 
probability current density $D\left ( \frac{\partial G_N({\bf{y}},t|{\bf{u}},0)}{\partial y_1}\right )_{y_1=0}$ over the hyperplane $\widetilde{\mathcal{W}}$ we get 
\begin{equation} 
 f_N(t|{\bf{u}}) = D \int_{\widetilde{\mathcal{W}}}dy_2 dy_3 \dots dy_N 
 \left ( \frac{\partial G_N({\bf{y}},t|{\bf{u}},0)}{\partial y_1}\right )_{y_1=0}
 ~~~~\text{for~process 2},\label{f-p2}
\end{equation}
where the Green's function is explicitly given in Eq.~(\ref{Green1}). 

To obtain the tail of the PDF $f_N(t|{\bf{u}})$, we need to find the large $t$ behavior of $G_N({\bf{y}},t|{\bf{u}},0)$, 
which can be obtained by expanding the function $\Phi_{{\bf{q}}}^{(N)}\left(\frac{{\bf{u}}}{\sqrt{t}} \right)$ in Eq. (\ref{phi}) 
for large $t$ and finite ${\bf u}$. One can show \cite{Kobayashi} that for large $t$ and finite ${\bf u}$,   
\begin{eqnarray}
 &&\Phi_{{\bf{q}}}^{(N)}\left(\frac{{\bf{u}}}{\sqrt{2Dt}} \right) =\det \limits_{1\leq i,j \leq N}\sin\left[q_i\frac{u_j}{\sqrt{2Dt}}\right] 
 \approx \gamma_N ~\mathcal{Y}_N({\bf{q}})
 ~\mathcal{Y}_N({\bf{u}})~\frac{1}{\left(\sqrt{2Dt}~\right)^{N^2}} +\mathcal{O}\left( t^{-(N^2+1)/2}\right), \label{Phi-expnd} \\
&&~~~~~~~~~~\text{where},~~~~\gamma_N=\frac{(-1)^{\frac{N(N-1)}{2}}}{\prod \limits_{i=1}^N(2i-1)!} 
~~~~~~~\text{and}~~~~~~~\mathcal{Y}_N({\bf{u}}) = \prod_{i=1}^{N}u_i\;\prod_{1\leq i<j \leq N}(u_j^2-u_i^2). \label{mathcalY}
\end{eqnarray}  
Plugging the above large $t$ approximation (\ref{Phi-expnd}) into Eq. (\ref{Green1}) and performing the integrations over the variables $q_i$, we get 
\begin{equation}
  G_N({\bf{y}},t|{\bf{u}},0) \simeq \frac{\mathcal{Y}_N({\bf u})}{(\sqrt{2Dt})^{N^2}}~\left 
  [\frac{\left(\frac{2}{\pi} \right)^{\frac{N}{2}}}{\prod \limits_{i=1}^{N}(2i-1)!}
  ~\frac{1}{(\sqrt{2Dt})^N}~\text{exp}\left[-\frac{{\bf y}^2}{4Dt} \right]~\mathcal{Y}_N\left( \frac{{\bf y}}{\sqrt{2Dt}}\right) \right].
  \label{Green2}
\end{equation}
Finally, putting this large $t$ form of $G_N({\bf{y}},t|{\bf{u}},0)$ into Eqs.~(\ref{f-p1}) and (\ref{f-p2}) 
and 
performing the rest of the integrations over the variables  $y_i$, we obtain 
\begin{equation}
 f_N(t_s|{\bf{u}}) \sim D\delta_N\frac{\mathcal{Y}_N({\bf{u}})}{(Dt_s)^{\zeta}},\;\;\text{with}\;\zeta= \frac{N^2}{2}+1, \label{survival} 
\end{equation}
for both process 1 and process 2, where $\delta_N$ is an $N$ dependent constant different for process 1 and process 2. 
The explicit expressions of $\delta_N$ for both processes 
are given in Appendix \ref{delta_N}.
The above result for $f_N(t_s|{\bf{u}}) $ has also been proved in \cite{Krattenthaler,Bray04} for process 2.
Plugging the large $t_s$ behavior of $f_N(t_s|{\bf{u}})$ from Eq. (\ref{survival}) into Eqs.~(\ref{scaling1}) and (\ref{scaling2}) we get, 
\begin{equation}
 1-Q_N(L|{\bf{u}}) \propto B_N\mathcal{Y}_N({\bf u})/{L^{N^2}},~\text{for~large~}L \label{Q-frm-argu}
\end{equation}
where, the function $\mathcal{Y}_N({\bf u})$ is given in Eq. (\ref{mathcalY}) and $B_N$ is an $N$ dependent constant. 
Upon deriving both sides of the Eq. (\ref{Q-frm-argu}) with respect to $L$ at $L=m$, we get the tail of the PDF 
$p_N(m|{\bf u})$ as given in Eq. (\ref{P_N_multi}) with $\nu_N=N^2+1$ and $\mathcal{A}_N = N^2B_N\mathcal{Y}_N({\bf u})$.
This heuristic argument also provides a rough estimate of $B_N$ for large $N$ and that is 
$B_N \sim \text{exp}\left(\frac{N^2}{2} \log N \right)$. Similar scaling arguments have been successfully used to study 
the distribution of the global maximum $m_N$ of $N$ non-interacting particles till their first exit from the half space \cite{KrapivskySatya2010}. 

This result (\ref{Q-frm-argu}) is in line with the following general results valid for a generic self-affine process. 
For such a process $x(t)$ starting from $x>0$, the cumulative distribution $Q(L|x)=\text{Prob.}(m\leq L|x)$ of the maximum $m$ 
till the stopping time $t_s$ (time of first passage through $x=0$), or equivalently the exit probability $Q(L|x)$ from the box $[0, L]$ through the origin, 
has been recently studied in \cite{Satyarosso10} where it was shown that $1-Q(L|x) \sim (x/L)^{\phi}$ in the $(x/L) \to 0$ limit. 
The exponent $\phi$ is related to the persistence exponent 
$\theta_p$ and the Hurst exponent $H$ via the scaling relation $\phi=\theta_p/H$ \cite{Satyarosso10}. The persistence exponent $\theta_p$ characterizes the 
late time power-law decay of the survival probability, i.e. the probability that the process stays on the positive half-axis up to time $t$ 
\cite{Bray13,Satyacurrsci99}, 
whereas the Hurst exponent characterizes the typical growth of $x(t) \sim t^H$ with time $t$. Thus, the PDF of the maximum
decays for large $m$ as $P(m|x) \sim m^{-\phi-1}$ with $\phi = \theta_p/H$. From Eq. (\ref{survival}) we see that for the multiparticle 
process the corresponding persistence exponent is $\theta_p = \frac{N^2}{2}$. If we consider this $N$-particle process as a single self-affine process in $N$-dimensional 
space with $H=1/2$, then the general argument from \cite{Satyarosso10} suggests that $1-Q(L|{\bf x}) \sim (1/L)^{\phi}$, with $\phi=(\theta_p/H)=N^2$, which is in 
accordance with Eq. (\ref{Q-frm-argu}), although this argument can not predict the precise dependence on the initial positions. 
In the next subsection we prove $\nu_N = N^2+1$ on firmer grounds and compute the amplitude exactly.



\subsection{The distribution $p_N(m|{\bf{u}})$ of the global maximum for $N>2$}
\label{morethan2}
\noindent 
Here we study the distribution $p_N(m|{\bf{u}})$ of the global maximum $m_N$ of $N$ identical (i.e. with identical diffusion constant 
$D_1=D_2=...=D_N=D$) vicious walkers using the $N$-particle Green's function. 
We first compute the cumulative probability 
$Q_N(L|{\bf{u}})=\text{Prob.}(m_N \leq L | {\bf{u}})$ which 
represents the probability that the global maximum $m_N$ of the rightmost particle is less or equal to $L$ 
given that the walkers, starting from positions ${\bf{u}}=(u_1,u_2,...,u_N)$ stayed non-intersecting till the ``stopping time'' $t_s$. 
Upon taking the derivative of $Q_N(L|{\bf{u}})$ with respect to $L$ at $L=m$ we get the PDF $p_N(m|{\bf{u}})$.  
To compute this cumulative probability $Q_N(L|{\bf{u}})$ we consider the first exit problem of a single $N$-dimensional 
Brownian walker ${\bf{u(t)}}=(u_1(t),u_2(t),...,u_N(t))$ from the region $\mathcal{T}_N(L)= \{ {\bf{u}} \in {\mathbb{R}}_+^N | 0<u_1<u_2<...<u_N<L\}$, 
as done for the $N=2$-particle case (see Fig. \ref{fig2}). For process 1 we consider the first exit probability of the walker 
through any of the boundaries $u_1=0$ or $u_{i+1}=u_{i}$ with $i=1,2,...,N-1$. 
These exit events correspond, in the original $N$-particle problem,
to the following events: (a) leftmost particle crossing the origin for the first time before any two particles meet or (b) any two particle meet for the first 
time before the leftmost particle hits the origin. 
On the other hand for process 2, we consider the first exit probability of the $N$-dimensional walker only 
through the boundary $u_1=0$. This event corresponds to the leftmost particle hitting the origin for the first time before any two particles meet in the $N$-particle 
picture. We denote this first exit probability for both process 1 and process 2 by $F_N({\bf u},L)$.  
For process 1, one can see that $F_N({\bf u},L)$ is equal to the 
time integration from $t=0$ to $t=\infty$ of the total outward probability flux through all the boundaries of $\mathcal{T}_N(L)$ (which is equal to 1) 
minus time integration of the outward flux through the boundary $u_N=L$ whereas for process 2 $F_N({\bf u},L)$ is equal to the time integration  
of the outward flux only through the 
boundary $u_1=0$. Hence, the probability $F_N({\bf u},L)$ can be expressed in terms of the Green's function $G^{(L)}_N({\bf y},t~|~{\bf u},0)$ as, 
\begin{eqnarray}
 F_N({\bf u},L) &=& 
1 +    D\int \limits_0^{\infty}dt\int_{W_0^L} d^{N-1}{\bf{y}}
\left(\frac{\partial G_N^{(L)}}{\partial y_N}\right)_{y_N=L}~~~~\text{for~process~1} \label{Exitprob-N-1} \\ 
\text{and}~~~F_N({\bf u},L) &=&~~~~~  D\int \limits_0^{\infty}dt\int_{W_0^L} d^{N-1}{\bf{y}}
\left(\frac{\partial G_N^{(L)}}{\partial y_1}\right)_{y_1=0}~~~~\;\,\text{for~process~2}, \label{Exitprob-N-2}
 \end{eqnarray}
where, we have introduced the notations 
 \begin{eqnarray} 
 \int_{W_a^b} d^{N-1}{\bf{y}} &=&
\int \limits_a^b dy_{N-1}\int \limits_a^{y_{N-1}} dy_{N-2}...\int \limits_a^{y_2} dy_1 ~~~~~\text{for~process~1},  \label{notation1} \\
\text{and}~~~\int_{W_a^b} d^{N-1}{\bf{y}} &=&
\int \limits_a^b dy_{N}\int \limits_a^{y_{N}} dy_{N-1}...\int \limits_a^{y_3} dy_2  ~~~~~~~~~~~\text{for~process~2}. \label{notation2} 
\end{eqnarray}
The Green's function used in Eqs. (\ref{Exitprob-N-1}) and (\ref{Exitprob-N-2}) represents the probability density that $N$ non-intersecting Brownian walkers, 
starting initially from ${\bf u}$, reach ${\bf y}$ in time $t$ . From Karlin-McGregor formula \cite{karlin}, 
it can be written in terms of a determinant of single particle propagators inside a box $[0,L]$ as,   
\begin{eqnarray}
 &&G^{(L)}_N({\bf y},t~|~{\bf u},0) = \det \limits_{1 \leq i,j \leq N} 
 \left [\sum \limits_{m=-\infty}^{\infty} g(u_i,y_j+2mL,t)\right], \label{Green00} 
\end{eqnarray}
where the function $g(u,y,t)$ is given in Eq. (\ref{gsingle}). 
Putting this form of the Green's function in Eqs. (\ref{Exitprob-N-1}) and (\ref{Exitprob-N-2}), one can see that $F_N({\bf u},L)$ can be expressed for both process 1 
and process 2 as 
\begin{equation}
 F_N({\bf u},L) = \sum \limits_{\{{\bf{m}}\}} \mathcal{J}_{{\bf m}}^{(N)}({\bf u},L)
 ~~\text{where}~~~
 \sum \limits_{\{{\bf{m}}\}} \equiv \sum\limits_{m_1=-\infty}^{\infty} \sum\limits_{m_2=-\infty}^{\infty}...\sum\limits_{m_N=-\infty}^{\infty}\;, \label{Exitprob-series}
 \end{equation}
 and
 \begin{eqnarray}
&&\hspace*{-1.cm}  \mathcal{J}_{{\bf m}}^{(N)}({\bf u},L) = \delta_{{\bf m,0}} + D\int \limits_0^{\infty}dt\int_{W_0^L} d^{N-1}{\bf{y}}
\left(\frac{\partial }{\partial y_N}\det \limits_{1 \leq i,j \leq N}g(u_i,y_j+2m_jL,t) \right)_{y_N=L}\text{for~process 1} \label{mcalJP1}\\
&&\hspace*{-1.cm}\mathcal{J}_{{\bf m}}^{(N)}({\bf u},L) = ~~~~~~~~~~D\int \limits_0^{\infty}dt\int_{W_0^L} d^{N-1}{\bf{y}}
\left(\frac{\partial }{\partial y_1} \det \limits_{1 \leq i,j \leq N}g(u_i,y_j+2m_jL,t) \right)_{y_1=0}\;\;\text{for~process 2}\;. \label{mcalJP2}
 \end{eqnarray}
We will see later that the above form of $F_N({\bf u},L)$ in Eq. (\ref{Exitprob-series}) will be convenient to compute the large $L$ asymptotics which will be needed to compute the tail of 
the PDF $p_N(m|{\bf u})$ [see Eq. (\ref{P_N_multi})].
Once we know $F_N({\bf u},L)$, the cumulative probability $Q_N(L|{\bf{u}})$ is obtained from the ratio (as done in the $N=2$-particle case)
\begin{equation}
 Q_N(L|{\bf{u}})= \frac{F({\bf{u}},L)}{S_N({\bf u})}, \;\;\;
 \text{where}~~S_N({\bf u})=\lim \limits_{L \to \infty} F({\bf{u}},L)\;.\label{P0_N}
\end{equation}
This ratio represents the fraction of such group of $N$ Brownian trajectories 
starting from positions ${\bf u}=(u_1,u_2,...,u_N)$, which have 
global maximum $m_N \leq L$ and stay mutually non-intersecting till the stopping time $t_s$. In the denominator, 
the quantity $S_N({\bf u})$ in Eq. (\ref{P0_N}) represents the probability that the process will ``stop'' ultimately. Clearly, for process 1 this 
probability is exactly one whereas for process 2 this probability is smaller than one ($S_N({\bf u})< 1$) and expressed as 
\begin{eqnarray}
S_N({\bf u})=D\frac{1}{N!} \left(\frac{2}{\pi}\right)^N\int_0^{\infty}d\tau~\int_{W_0^{\infty}} d^{N-1}{\bf{y}}~\int_0^{\infty}d^N{\bf{k}}~
~\text{exp}\left[-{\bf{k}}^2 D\tau\right ]~
\Phi_{{\bf{k}}}^{(N)}\left({\bf u}\right) \left\{\frac{\partial}{\partial y_1}
\Phi_{{\bf k }}^{(N)}\left({\bf{y}} \right)\right\}_{y_1=0}\label{S_N-explicit} 
\end{eqnarray}
where $\Phi_{{\bf{k}}}^{(N)}\left({\bf u}\right)$ is given in Eq. (\ref{phi}). 
A more explicit expression of $S_N({\bf u})$ is given in Eq. (\ref{S_N}).
One can, in principle, compute the integral in Eq. (\ref{S_N-explicit}) for any given $N$ and ${\bf u} \equiv (u_1,u_2,....,u_N)$. For $N=2$, the 
probability $S_2({\bf u})$ is explicitly given by $S_2({\bf u})= \frac{4}{\pi}\arctan \left(\frac{u_2}{u_1} \right) - 1$ [see Eq. (\ref{exit-ult-2part}) for $D_1=D_2$]. For $N=3$, an explicit expression of 
$S_3({\bf u})$ is given in Eq. (\ref{S_3}). 

To find the large $m$ form of the distribution $p_N(m|{\bf{u}})$,
we first look at the large $L$ limit of $F({\bf u}, L)$ to get the large $L$ form of $Q_N(L|{\bf{u}})$ from Eq. (\ref{P0_N}). 
We show below in Eq. (\ref{F3}) that, for both process 1 and process 2 the probability $F({\bf u}, L)$ has the following large $L$ form 
\begin{eqnarray}
F({\bf u}, L) &=&  S_N({\bf u}) - B_N \frac{\mathcal{Y}_N({\bf{u}})}{L^{N^2}} + \mathcal{O}\left( \frac{1}{L^{N^2+1}}\right)~~\text{where,} \label{F0}\\
\mathcal{Y}_N({\bf u}) &=& \prod_{i=1}^{N}u_i\;\prod_{1\leq i<j \leq N}(u_j^2-u_i^2), \label{mathcalY1}
\end{eqnarray}
and $B_N$ is an $N$ dependent constant. 
Hence from the ratio in Eq. (\ref{P0_N}) we get 
\begin{equation}
  Q_N(L|{\bf{u}})= 1- B_N \frac{\mathcal{Y}_N({\bf{u}})}{S_N({\bf u})}~\frac{1}{L^{N^2}}+ \mathcal{O}\left( \frac{1}{L^{N^2+1}}\right),
\end{equation}
from which we finally obtain 
\begin{equation}
p_N(m|{\bf u})= \left(\frac{\partial Q_N(L|{\bf{u}})}{\partial L} \right)_{L=m} \simeq  
N^2B_N \frac{\mathcal{Y}_N({\bf{u}})}{S_N({\bf u})}~\frac{1}{m^{(N^2+1)}}~~\text{for}~~m \gg u_N, \label{dist-global-max}
\end{equation}
as announced in Eq. (\ref{P_N_multi}). This asymptotic result indicates that for $N$ walkers, integer moments of $m_N$ up to order
$(N^2 - 1)$ are finite, while higher integer moments are infinite. Therefore as $N$ increases, the distribution becomes narrower and narrower as 
expected but this happens in a nontrivial way. It is instructive to compare the prefactor of the algebraic tail of $p_N(m|{\bf u})$ in Eq. (\ref{dist-global-max}) 
with the same amplitude in the non-interacting 
case in Eq. (\ref{non_int}). Besides the factor  $\prod_i u_i$ which is in common with the non-interacting case, the non-intersecting condition is encoded in this amplitude (\ref{dist-global-max}) through the Vandermonde determinant in $\mathcal{Y}_N({\bf u})$ (\ref{mathcalY1}). The appearance of the Vandermonde determinant is 
reminiscent of the connection between the present vicious walkers problem and random matrix theory~\cite{Schehr08}. 

In the following we give an outline of the proof of Eqs. (\ref{F0}) and (\ref{mathcalY1}) for process 2. 
For process 1 one can follow similar calculations starting from Eqs. (\ref{Exitprob-series}) and (\ref{mcalJP1})  
to arrive at Eq. (\ref{F0}). We start by using the following identity
\begin{equation}
 g(u,y,t)=\frac{1}{\sqrt{4\pi Dt}}\left[ \text{exp}\left(-\frac{(u-y)^2}{4Dt} \right) 
 - \text{exp}\left(-\frac{(u+y)^2}{4Dt}\right)\right] = \frac{2}{\pi} \int_0^{\infty}dk~\sin(ku)\sin(ky)~e^{-k^2Dt}, \label{identity}
\end{equation}
in the expression of the Green's function $g(u,y,t)$ in Eq. (\ref{mcalJP2}). By performing then some algebraic manipulations, one can show from 
Eq. (\ref{Exitprob-series}) that 
the first exit probability $F({\bf u}, L)$ can be written in the following form 
\begin{eqnarray}
&&F({\bf u}, L)=S_N({\bf u}) + \frac{1}{N!} \left(\frac{2}{\pi}\right)^N 
 \sum \limits_{\{{\bf{m}}\}} q_N^{({\bf{m}})}({\bf u},L)~~~\text{where}, \label{F2} \\
 &&q_N^{({\bf{0}})}({\bf u},L) = -D\int_0^{\infty}d\tau~\int_{W_1^{\infty}} d^{N-1}{\bf{z}}~\int_0^{\infty}d^N{\bf{q}}~
~\text{exp}\left[-{\bf{q}}^2 D\tau\right ]~
\Phi_{{\bf{q}}}^{(N)}\left(\frac{{\bf{u}}}{L} \right) \left\{\frac{\partial}{\partial z_1}
\Phi_{{\bf{q}}}^{(N)}\left({\bf{z}} \right)\right\}_{z_1=0},\label{qm0} \\
&&q_N^{({\bf{m}})}({\bf u},L) = D\int_0^{\infty}d\tau~\int_{W_0^1} d^{N-1}{\bf{z}}~\int_0^{\infty}d^N{\bf{q}}~
~\text{exp}\left[-{\bf q}^2 D\tau \right ]~
\Phi_{{\bf{q}}}^{(N)}\left(\frac{{\bf{u}}}{L} \right) \left\{\frac{\partial}{\partial z_1}
\Phi_{{\bf{q}}}^{(N)}\left({\bf{z}+2{\bf m}} \right)\right\}_{z_1=0}
\label{qm}
\end{eqnarray}
and $S_N({\bf u})$ and $\Phi_{{\bf{k}}}^{(N)}\left({\bf u}\right)$ are given in Eqs. (\ref{S_N-explicit}) and (\ref{phi}) respectively.  
To arrive at the above expression of $F({\bf u}, L)$ we used that
$\int_{W_0^{L}}d^{N-1}y=\int_{W_0^{\infty}}d^{N-1}y -\int_{W_{L}^{\infty}}d^{N-1}y$ and performed the following change of variables 
${\bf k}=\frac{{\bf q}}{L}$, ${\bf y}={\bf z}~L$ 
inside the integrations. 
Note that this expression of $F({\bf u}, L)$ given in Eq. (\ref{F2}) 
is more suitable for obtaining a large $L$ asymptotic as 
the $L$ dependence is contained in the function $\Phi_{{\bf{q}}}^{(N)}\left(\frac{{\bf{u}}}{L} \right)$ which is a 
determinant of $\sin\left[q_i\frac{u_j}{L}\right]$ (\ref{phi}). The 
large $L$ expansion of $\Phi_{{\bf{q}}}^{(N)}\left(\frac{{\bf{u}}}{L} \right)$ is obtained from Eq. (\ref{Phi-expnd}) by replacing $t=\frac{L^2}{2D}$. Plugging this 
large $L$ behavior in Eqs. (\ref{qm0}) and (\ref{qm}),
we get from Eq. (\ref{F2}) 
\begin{equation}
 F({\bf u}, L) =  S_N({\bf u}) - B_N \frac{\mathcal{Y}_N({\bf{u}})}{L^{N^2}}
+ \mathcal{O}\left( \frac{1}{L^{N^2+1}}\right)\;, \; L \to \infty \label{F3} \\
\end{equation}
where, 
\begin{eqnarray}
&&~~~~
B_N=r_N \left[ R_N({\bf{0}})-\sum \limits_{\{{\bf{m}\neq {\bf 0}}\}} R_N({\bf{m}}) \right],~~\text{with}~
r_N= \frac{(-1)^{\frac{N(N-1)}{2}}}{2N!\prod \limits_{i=1}^N(2i-1)!} \left(\frac{2}{\pi}\right)^{N} \;, \; {\rm and}
\label{F31} \\
&& ~~~~~R_N({\bf{0}})=\int_0^{\infty}d\tau~\int_{W_1^{\infty}} d^{N-1}{\bf{z}}~\int_0^{\infty}d^N{\bf{q}}~
~\text{exp}\left[- \frac{{\bf q}^2\tau}{2}\right ]~\mathcal{Y}({\bf q})~
 \left\{\frac{\partial}{\partial z_1}\Phi_{{\bf{q}}}^{(N)}\left({\bf{z}} \right)\right\}_{z_1=0}, \label{B_N_1} \\
&& ~~~~~R_N({\bf{m}})=\int_0^{\infty}d\tau~\int_{W_0^1} d^{N-1}{\bf{z}}~\int_0^{\infty}d^N{\bf{q}}~
~\text{exp}\left[- \frac{{\bf q}^2 \tau}{2} \right ]~\mathcal{Y}({\bf q})~
 \left\{\frac{\partial}{\partial z_1}\Phi_{{\bf{q}}}^{(N)}\left({\bf{z}+2{\bf m}} \right)\right\}_{z_1=0}, \label{B_N_2}
\end{eqnarray}
and the function $\mathcal{Y}({\bf q})$ is given in Eq. (\ref{mathcalY1}). The integration over the variables $z_i$ in the above formula 
is understood in terms of the notations given in Eq. (\ref{notation2}). Moreover one should note that the domain of integration 
corresponding to the integration over the variables $z_i$ in the case ${\bf m}={\bf 0}$ is different from the case 
${\bf m} \neq {\bf 0}$. 
Performing the integrations over $q_i$'s in Eqs. (\ref{B_N_1}) and (\ref{B_N_2}), one can rewrite the constant $B_N$ given in Eq. (\ref{F3}) as 
\begin{eqnarray}
B_N&=&d_N \left[ h_N({\bf{0}})-\sum \limits_{\{{\bf{m}\neq {\bf 0}}\}} h_N({\bf{m}}) \right],~~\text{with}~~~
d_N= \frac{1}{2\prod \limits_{i=1}^N(2i-1)!} \left(\frac{2}{\pi}\right)^{\frac{N}{2}},~~~\text{and}~
\label{B_N-1} \\
h_N({\bf{0}})&=&  \int_0^{\infty}d\tau~\tau^{-\frac{N(N+1)}{2}}\int_{W_1^{\infty}} d^{N-1}{\bf{z}}~
  \text{exp}\left(-\frac{{\bf z}^2}{2\tau} \right)\left[\frac{\partial}{\partial z_1} 
 \mathcal{Y}_N \left(\frac{{\bf z}}{\sqrt{\tau}}\right)\right]_{z_1=0} \nonumber \\
h_N({\bf{m}})&=&\int_0^{\infty}d\tau~{\tau^{-\frac{N(N+1)}{2}}}\int_{W_0^1} d^{N-1}{\bf{z}}~
 \left[\frac{\partial}{\partial z_1}~\text{exp}\left(-\frac{({\bf{z}}+2{\bf{m}})^2}{2\tau} \right) 
 \mathcal{Y}_N \left(\frac{{\bf{z}}+2{\bf{m}}}{\sqrt{\tau}}\right)\right]_{z_1=0}, \label{B_N-2} 
\end{eqnarray}
where, again, the function $\mathcal{Y}_N({\bf u})$ is given in Eq. (\ref{mathcalY1}). In principle one can compute numerically the constant $B_N$ given in Eq. (\ref{B_N-1}) 
for any given $N$. For instance, for $N=2$, such a numerical evaluation yields $B_2=2.25689\ldots$. Comparing the expression of $p_2(m|{\bf u})$ in Eq. (\ref{dist-global-max}) 
for $N=2$ with the Eq.~(\ref{p2-tail}), we see that the value of $B_2$ obtained from the Green's function method 
matches exactly with the value of $k_2/4 = \frac{3~\Gamma[1/4]^8}{20(4\pi)^3}=2.25689\ldots$ obtained previously in Eq. (\ref{k-consts}) for process 2 
through a completely different approach (using Schwarz-Christoffel mapping). 
For $N=3$ we obtain the numerical estimate from (\ref{B_N-1}) $B_3=16.3053..$. It is interesting to find the large $N$ asymptotic behavior of $B_N$. 
One can argue and we have checked it numerically with Mathematica that for large $N$ the dominant contribution to $B_N$ in Eq. (\ref{B_N-1}) comes from the term 
$h_N(0,0,..,0,-1)$.  Hence we conjecture that  
\begin{equation}
B_N \simeq d_N \Delta_N \,~~\text{for~large}~N,~~\text{where}~\Delta_N=-h_N(0,0,...,0,-1), \label{B_N_3}
\end{equation}
and $d_N$ is given in Eq. (\ref{B_N-1}).
From Eq. (\ref{B_N-2}), one can see that $\Delta_N$ has the following form  
\begin{equation}
\Delta_N=\int_0^{\infty}d\tau~{\tau^{-\frac{N^2+N+4}{2}}} \int_0^1dz_{N}~\text{exp}\left[-\frac{(z_{N}-2)^2}{2\tau} \right]
 ~(2-z_{N})^3~\mathcal{K}_{N-2}(z_{N},\tau), \label{h-const-0}
 \end{equation}
 where the explicit expression of the function $\mathcal{K}_{N}(x,\tau)$ is given in Eq. (\ref{mathcalK}) (see Appendix D). The large $N$ analysis of 
 $\mathcal{K}_{N}(x,\tau)$  can be carried out using analytical techniques from random matrix theory, namely Coulomb gas techniques \cite{Dyson,Dean08,Dean08II,Bohigas} (for a review see Ref. \cite{Majumdar14}). We then show in  
 Appendix \ref{B_N-lrg-N} that for large $N \gg 1$, the constant $\Delta_N$ grows as :
\begin{equation}
 \Delta_N = -h_N(0,0,...,0,-1) \approx \text{exp}\left(\frac{3N^2}{2} \log N \right).  \label{deltaN-lrgN}
\end{equation}
On the other hand, from the explicit expression of $d_N$ given in Eq. (\ref{B_N-1}) we get 
\begin{equation}
 d_N \approx \text{exp}\left(-\frac{N^2}{2} \log N \right),~~\text{for}~~N\gg 1 \;. \label{d_N-lrgN} 
\end{equation}
Hence using Eqs. (\ref{deltaN-lrgN}) and (\ref{d_N-lrgN}) in Eq. (\ref{B_N_3}) we finally get 
\begin{equation}
B_N \simeq  d_N \Delta_N \approx \text{exp}\left(\frac{N^2}{2} [\log N + o(\log N)]\right)~~\text{for}~~N\gg 1 \;, 
\end{equation}
where $o(\log N)$ represents terms smaller than $\log N$. This large $N$ asymptotic form of $B_N$ 
agrees with the rough estimate obtained from the heuristic argument in section \ref{argument}.

\section{Conclusion} \label{conclu}
To summarize, we have considered the extreme statistics of $N$ non-intersecting Brownian motions in one dimension (``vicious walkers''), till their survival. These  
Brownian particles ``survive'' over a random time interval $[0,t_s]$ where $t_s$ is usually called the ``stopping time''. 
We consider two different stopping mechanisms named process 1 and process 2 to define $t_s$. 
For process 1, the $N$-particle process gets ``stopped'' 
when either any of the two particles among $N$ particles meet each other for the first time before the leftmost one hits the origin or the 
leftmost particle hits the origin for the first time before any two particles meet each other (see Fig. \ref{fig1}). 
On the other hand, for process 2, the ``stopping time'' $t_s$ is determined 
from the first passage time of the leftmost walker given that no other two particles have met before $t_s$. 
For $N=2$ particles, we have computed exactly the joint 
cumulative distribution function $\mathcal{Q}(L_1,L_2|u_1,u_2)$ of the maxima of 
the leftmost and rightmost particle till their survival. This was done by solving a two-dimensional backward Fokker-Planck equation with the help 
of a conformal mapping, namely the Schwarz-Christoffel transformation. 
From the joint cumulative distribution we have obtained the marginal PDFs of the maxima of the first and second particle $p_1(m|u_1,u_2)$ 
and $p_2(m|u_1,u_2)$ respectively. 
For general $N$ identical walkers, we have computed the tail of the distribution 
$p_N(m|{\bf u})$ of the global maximum $m_N$ in two ways. The first one is using a heuristic argument based on the distribution $f_N(t|{\bf u})$ of the 
``stopping time'' while the second one is an exact calculation based on $N$-particle Green's function. 

This work raises several interesting questions, which certainly deserve further studies. The first extension of the present study is the computation of the  
exponent $\nu_N$ and the associated amplitude $\mathcal{A}_N(\bf{u},\bf{D})$ for $N>2$ particles and different diffusion constants. This is a challenging question from a technical point of view as, in this case, one can not use the Karlin-McGregor formula. In this paper, we have mostly focused on the distribution of the value of the global maximum $m_N$ of $N$ non-intersecting walkers till the stopping time $t_s$. Another interesting observable is not just
the actual value of the maximum $m_N$, but the time $t_m$ at which this maximum occurs before the
stopping time $t_s$. The PDF of $t_m$ was studied for vicious walkers over a fixed time interval \cite{Rambeau11}, with interesting application to stochastic growth processes \cite{Rambeau10}, and it will be interesting to study it for the stopped multi-particle process. Finally, another interesting open question concerns the distribution $p_1(m|{\bf u})$ of the maximum displacement $m_1$ of the leftmost walker for $N>2$. This is an interesting quantity 
as the maximal displacement $m_1$ travelled by the leftmost particle can be considered, for instance, as a measure of the common region \cite{Anupam13} visited by all the walkers.

\begin{acknowledgement}
We acknowledge support by ANR grant 2011-BS04-013-01 WALKMAT and in part by the Indo-French 
Centre for the Promotion of Advanced Research under Project~$4604-3$. G. S. also acknowledges support from the grant Labex PALM-RANDMAT.
\end{acknowledgement}

\appendix

\section{The constant $\delta_N$ in Eq. (\ref{survival})}
\label{delta_N}
\noindent
Here we give explicit and exact expression of the constant $\delta_N$ appearing in Eq. (\ref{survival}) for both processes 1 and 2. \\
For process 1 
\begin{eqnarray}
 \delta_N&=& \frac{\left(\frac{2}{\pi}\right)^{N/2}N^2}{2^{(\frac{N^2}{2}+1)}\prod \limits_{j=1}^{N} \Gamma[2i]} \int \limits_0^{\infty} dz_N 
 \int \limits_0^{z_N}dz_{N-1}...\int \limits_0^{z_3}dz_2 \int \limits_0^{z_2}dz_1 \; \text{exp}\left(-\sum \limits_{i=1}^N\frac{z_i^2}{2} \right)
 \prod \limits_{i=1}^N z_i \prod_{1\leq i<j\leq N}(z_j^2-z_i^2) \nonumber \\
 &=&~~~~~~~~~~~~\left(\frac{2}{\pi}\right)^N\frac{N}{2(N-1)!}~\frac{\prod \limits_{j=1}^{N} \Gamma \left[1+\frac{j}{2}\right]
\Gamma \left[1+\frac{j-1}{2}\right]}{\prod \limits_{i=1}^N\Gamma[2i]},~~~~\text{and}
\end{eqnarray}
for process 2 
\begin{eqnarray}
 \delta_N&=& \frac{\left(\frac{2}{\pi}\right)^{N/2}}{2^{(\frac{N^2}{2}+1)}\prod \limits_{j=1}^{N} \Gamma[2i]} \int \limits_0^{\infty} dz_N 
 \int \limits_0^{z_N}dz_{N-1}...\int \limits_0^{z_3}dz_2 \text{exp}\left(-\sum \limits_{i=2}^N\frac{z_i^2}{2} \right)
 \prod \limits_{i=2}^N z_i^3 \prod_{2\leq i<j\leq N}(z_j^2-z_i^2) \nonumber \\
 &=&~~~~~~~~~~~~\left(\frac{2}{\pi}\right)^N \frac{ N\sqrt{\pi}}{8~N!} 
~\frac{\prod \limits_{j=1}^{N-1} \Gamma \left[1+\frac{j}{2}\right]\Gamma \left[2+\frac{j-1}{2}\right]}{\prod \limits_{i=1}^N\Gamma[2i]}
\end{eqnarray}
where $\Gamma[x]$ is the Gamma function. These two expressions are obtained using exact formulas for Selberg integrals~\cite{Mehta}.

\section{Explicit expressions of the functions $\mathcal{B}(u_1,u_2,\alpha, \mu)$ and $\mathcal{A}_i(u_1,u_2, \mu)$. }
\label{expreuneqD}

Following the method explained in section \ref{section3}, one can compute the joint cumulative distribution function of $m_1$ and $m_2$ till
the stopping time $t_s$ for any diffusion coefficients $D_1, D_2$. We will not give the details here but only quote the results for the asymptotic
behaviors which one can extract from this exact calculation. One finds indeed
\begin{equation}
  \mathcal{Q}(L,\alpha L|u_1,u_2) \approx 1 - \frac{\mathcal{B}(u_1,u_2,\alpha, \mu)}{L^{\mu}} ~~\text{with}~~
  \mu=\frac{2\pi}{\pi-2\arctan \left( \sqrt{\frac{D_1}{D_2}} \right)}~~~\text{for}~~L\gg u_2
\end{equation} 

\begin{eqnarray}
 \mathcal{B}(u_1,u_2,\alpha,\mu) &=& \frac{1}{\pi}\mathcal{X}_{\theta}(a)~\left(\sqrt{\frac{D_2u_1^2+D_1u_2^2}{D_2}}\right)^{\frac{1}{\theta}}~
 \sin \left( \frac{\psi-\beta}{\theta}\right)
 ~~~~\text{for process 1 and }\label{BP1} \\
 \mathcal{B}(u_1,u_2,\alpha,\mu) &=& \frac{\theta}{\psi-\beta}\left(\frac{2a\theta+a-1}{2a(1+\theta)}\right)\mathcal{X}_{\theta}(a)
 ~\left(\sqrt{\frac{D_2u_1^2+D_1u_2^2}{D_2}}\right)^{\frac{1}{\theta}}~\sin \left( \frac{\psi-\beta}{\theta}\right)~~~~\text{for process 2}
 \label{BP2} \\
 \text{where},~\mathcal{X}_{\theta}(a)&=&\left (\theta~h_{\theta}(a)\sqrt{a} \right)^{\frac{1}{\theta}},  
 ~\psi=\arctan\left( \frac{\sqrt{D_1}u_2}{\sqrt{D_2}u_1}\right),~~
\theta=\frac{1}{2}-\frac{\beta}{\pi}~~\text{with}~~\beta=\arctan \left( \sqrt{\frac{D_1}{D_2}}\right), 
\end{eqnarray}
which yields the asymptotic behaviors of the PDFs:
\begin{equation}
p_i(m|u_1,u_2) \approx \frac{\mathcal{A}_i(u_1,u_2,\mu)}{m^{\nu}} ~~\text{with}~~\nu=\mu+1~~~\text{for}~~m\gg u_2 \;~\text{and}~i=1,2. 
\end{equation}
The amplitudes $\mathcal{A}_i(u_1,u_2,\mu)$ are explicitly given by
\begin{eqnarray}
  \mathcal{A}_1(u_1,u_2, \mu) &=& \frac{1}{\pi\theta} 
 \left( \frac{\pi\theta\sqrt{D_2u_1^2+D_1u_2^2}}{\sqrt{D_2}}\right)^{\frac{1}{\theta}}
 ~\sin \left( \left[\arctan \left(\frac{\sqrt{D_1}u_2}{\sqrt{D_2}u_1}\right)-\beta \right ]\theta^{-1}\right)~~{\text{process 1}} \label{AaP11} \\
 \mathcal{A}_1(u_1,u_2, \mu) &=& 
 \left(\frac{\theta}{1+\theta}\right)\left( \frac{\pi\theta
 \sqrt{D_2u_1^2+D_1u_2^2}}{\sqrt{D_2}}\right)^{\frac{1}{\theta}}
 \frac{\sin \left(  \left[\arctan \left(\frac{\sqrt{D_1}u_2}{\sqrt{D_2}u_1}\right)-\beta \right ]\theta^{-1}\right)}
 {\left[\arctan \left(\frac{\sqrt{D_1}u_2}{\sqrt{D_2}u_1}\right)-\beta\right ]}~~{\text{process 2}} \;,
 ~ \label{AaP12} 
\end{eqnarray}
together with
\begin{eqnarray}
 \mathcal{A}_2(u_1,u_2, \mu) &=& \frac{1}{\pi\theta} 
 \left( \frac{\theta~\Gamma[\theta]\Gamma[\frac{1}{2}-\theta]\sqrt{D_2u_1^2+D_1u_2^2}}{\sqrt{\pi}\sqrt{D_1+D_2}}\right)^{\frac{1}{\theta}}
 ~\sin \left( \left[\arctan \left(\frac{\sqrt{D_1}u_2}{\sqrt{D_2}u_1}\right)-\beta \right ]\theta^{-1}\right)\,{\text{process 1}} \label{AaP21} \\
 \mathcal{A}_2(u_1,u_2, \mu) &=& 
 \left(\frac{1+2\theta}{2(1+\theta)}\right)\left( \frac{\theta~\Gamma[\theta]\Gamma[\frac{1}{2}-\theta]
 \sqrt{D_2u_1^2+D_1u_2^2}}{\sqrt{\pi}\sqrt{D_1+D_2}}\right)^{\frac{1}{\theta}}
 \frac{\sin \left(  \left[\arctan \left(\frac{\sqrt{D_1}u_2}{\sqrt{D_2}u_1}\right)-\beta \right ]\theta^{-1}\right)}
 {\left[\arctan \left(\frac{\sqrt{D_1}u_2}{\sqrt{D_2}u_1}\right)-\beta\right ]}\,{\text{process 2}} \;. \nonumber\\
 ~ \label{AaP22} 
\end{eqnarray}
\section{The exit probability $S_N({\bf u})$ given in Eq. (\ref{S_N-explicit})}
\label{Psi-function}
After performing the integrations over $k_i$ variables in Eq. (\ref{S_N-explicit}), the exit probability $S_N({\bf u})$ can be rewritten as 
\begin{equation} 
 S_N({\bf u}) = \frac{4}{\sqrt{\pi}}\sum \limits_{i=1}^N (-1)^{i+1}\int \limits_0^{\infty} dt~\frac{u_i}{\sqrt{4t}}~\text{exp}\left( - \frac{u_i^2}{4t}\right)
\int \limits_0^{\infty} dz_{N}\int \limits_0^{z_{N}} dz_{N-1}...\int \limits_0^{z_3} dz_2~~ 
\det \left[ \bm{\tilde{\Lambda}}_i\left(\frac{{\bf u}}{\sqrt{4t}};z_2,z_3,...,z_N \right) \right] \;, \label{S_N} 
\end{equation}
where $\bm{\tilde{\Lambda}}_i$ is an $(N-1)\times(N-1)$ matrix obtained by removing the $i^{\text{th}}$ row and $1^{\text{st}}$ column from the $N\times N$ matrix 
$\bm{\tilde{\Lambda}}$ whose elements are given by  
$[\bm{\tilde{\Lambda}}]_{i,j} =\sqrt{\pi} g\left(\frac{u_i}{\sqrt{4t}},z_j,\frac{1}{4D} \right)$. We recall that the function 
$g\left(u,y,t\right)$ is given by
\begin{equation}
 g(u,y,t)=\frac{1}{\sqrt{4\pi Dt}}\left(\text{exp}\left[-\frac{(y-u)^2}{4Dt}\right] -\text{exp}\left[-\frac{(y+u)^2}{4Dt}\right]\right). 
\end{equation}
For $N=3$ we can obtain an explicit expression for the exit probability $S_3({\bf u})$. It is given by 
\begin{eqnarray}
 S_3(u_1,u_2,u_3)= &&\left \{\Psi(u_1,u_2,u_3) - \Psi(u_1,u_3,u_2) \right\} +  
 \left \{\Psi(u_2,u_3,u_1) - \Psi(u_2,u_1,u_3) \right\} \nonumber \\
 &&~~~~~~~~~~~~~~~ + \left \{\Psi(u_3,u_1,u_2) - \Psi(u_3,u_2,u_1) \right\} \label{S_3}
\end{eqnarray}
where
\begin{eqnarray}
 &&\Psi(x,y,z)= \frac{x}{\pi} \Bigg[ \\
 &&\text{ArcTan}\left(\frac{ x \left(\sqrt{x^2+y^2+z^2}+y-z\right) }{ x^2+\sqrt{\left(2 x^2+(y+z)^2\right) \left(2 z
   \left(z-\sqrt{x^2+y^2+z^2}\right)+x^2+y^2\right)}-(y+z)\left(\sqrt{x^2+y^2+z^2}-z\right) }  \right) \nonumber \\
   && -\text{ArcTan}\left(\frac{x \left(\sqrt{x^2+y^2+z^2}-y-z\right)}{ x^2+\sqrt{\left(2 x^2+(y+z)^2\right) \left(2 z
   \left(z-\sqrt{x^2+y^2+z^2}\right)+x^2+y^2\right)}+(y+z)\left(\sqrt{x^2+y^2+z^2}-z\right) }\right) \nonumber \allowdisplaybreaks[4] \\
 && -\text{ArcTan}\left(\frac{x \left(\sqrt{x^2+y^2+z^2}+y+z\right)}{ x^2+\sqrt{\left(2 x^2+(y+z)^2\right) \left(2 z
   \left(z-\sqrt{x^2+y^2+z^2}\right)+x^2+y^2\right)}-(y+z)\left(\sqrt{x^2+y^2+z^2}-z\right) }\right) \nonumber \\
&& +\text{ArcTan}\left(\frac{x \left(\sqrt{x^2+y^2+z^2}-y+z\right)}{ x^2+\sqrt{\left(2 x^2+(y+z)^2\right) \left(2 z
   \left(z-\sqrt{x^2+y^2+z^2}\right)+x^2+y^2\right)}+(y+z)\left(\sqrt{x^2+y^2+z^2}-z\right) }\right) \Bigg ]. \nonumber
\end{eqnarray}
For larger values of $N$ is does not seem possible to express the integrals in Eq. (\ref{S_N}) in terms of simple elementary functions.

\section{Large $N$ asymptotic of the constant $\Delta_N$ given in Eq. (\ref{h-const-0})}
\label{B_N-lrg-N}
Here we find the large $N$ asymptotic of the constant $\Delta_N$ in Eq. (\ref{h-const-0}). We rewrite it as
\begin{eqnarray}
&&\hspace*{-0.5cm}\Delta_N=\int_0^{\infty}d\tau~{\tau^{-\frac{N^2+N+4}{2}}} \int_0^1dz_{N}~\text{exp}\left[-\frac{(z_{N}-2)^2}{2\tau} \right]
 ~(2-z_{N})^3~\mathcal{K}_{N-2}(z_{N},\tau)~~~\text{where}, \label{h-const} \\
 &&\hspace*{-0.5cm}\mathcal{K}_{N-2}(z_{N},\tau)=
 \int_{W_0^{z_{N}}} d^{N-2}{\bf{z}}~
  \text{exp}\left[-\sum \limits_{i=2}^{N-1}\frac{z_i^2}{2\tau} \right] ~\prod \limits_{i=2}^{N-1}\left(\frac{z_i}{\sqrt{\tau}}\right)^3 
  ~\prod \limits_{i=2}^{N-1}\left[\frac{(z_N-2)^2}{\tau}-\frac{z_i^2}{\tau} \right] \prod \limits_{2\leq i <j \leq N-1}
  \left(\frac{z_j^2}{\tau} - \frac{z_i^2}{\tau} \right), \nonumber \\~~~ \label{mathcalK} 
 \end{eqnarray}
where the notation $\int_{W_0^{z_{N}}}$ is explained in Eq. (\ref{notation2}). 
Using the following identity  
 \begin{equation}
   \int_{W_0^{z_{N}}} d^{N}{\bf{z}}~f({\bf z}) \prod \limits_{1\leq i <j \leq N}
  \left(\frac{z_j^2}{\tau} - \frac{z_i^2}{\tau} \right) = \frac{1}{N!} \int_0^{z_N}...\int_0^{z_N} dz_1dz_2...dz_{N}~ f({\bf z}) 
  \prod \limits_{1\leq i <j \leq N}
  \left|\frac{z_j^2}{\tau} - \frac{z_i^2}{\tau} \right| \;,
 \end{equation}
 valid for any well behaved function $f({\bf z})$, one can show that $\Delta_N$ can be expressed as 
 \begin{eqnarray}
 \Delta_N&=& \frac{1}{2^{N-2}(N-2)!} \int_0^{\infty}d\tau~{\tau^{-\frac{N^2+6}{2}}}\int_0^1dx~\text{exp}\left[-\frac{(x-2)^2}{2\tau} \right]
 ~(2-x)^3 ~ \mathcal{M}_{N-2}\left( \frac{x^2}{\tau},\frac{(2-x)^2}{\tau}\right) \;, \label{h1} \\
\text{where}&&\mathcal{M}_{M}\left(c,d\right) = \int_0^{c} dq_1...\int_0^{c} dq_M~ \text{exp}[-\mathcal{V}({\bf{q}})], \label{M-partition} \\
 \text{with} &&\mathcal{V}({\bf{q}}) = \frac{1}{2} \left[ \sum \limits_{i=1}^M [q_i - 2\log(q_i) - 2 \log(d-q_i)] - \sum \limits_{i\neq j}^M \log|q_j-q_i|\right]. 
 \label{Hamiltonian-1}
\end{eqnarray}
The above expression of $\mathcal{M}_{M}\left(c,d\right)$ in (\ref{M-partition}) can be interpreted as a 
partition function of $M$ particles with coordinates ${\bf{q}}=(q_1,q_2,...,q_M)$ which are subject to a global linear+logarithmic external potential and 
interacting via two dimensional repulsive Coulomb potential $\frac{1}{2}\sum \limits_{i\neq j}^M \log|q_j-q_i|$. 
Such systems of particles are generally known as Coulomb gas in the literature \cite{Dyson,Mehta,Forrester}. The expression in Eq. (\ref{M-partition}) indicate that here  
the particles are confined on a line segment $[0,c]$ by putting two infinite walls at $q=0$ and $q=c$. Similar Coulomb gas with walls appears in the study of the cumulative distribution of the largest eigenvalue of a $M\times M$ random Wishart matrix \cite{Bohigas}, where the eigenvalues are $\ge 0$ by construction (see Ref. \cite{MS14} for a recent review). The large $M$ analysis of $\mathcal{M}_{M}\left(c,d\right)$ can be performed using a saddle point method as done in \cite{Dean08,Dean08II,Bohigas}. Following 
Refs. \cite{Dean08,Dean08II,Bohigas}, we first define a density of particles as 
 $\rho(q) = \frac{1}{M}\sum \limits_{i=1}^M \delta(q-q_i)$ 
and then express the function $\mathcal{M}_{M}\left(c,d\right)$ in Eq. (\ref{M-partition}) in terms of $\rho(q)$. As a result the integral in Eq. (\ref{M-partition}) 
becomes a functional integral with exponential weight in the integrand. This saddle point calculation yields to leading order for large $M$:
\begin{equation}
\mathcal{M}_{M}\left(c,d\right) \approx \text{exp}\left[\frac{M^2}{2}\log(M) \right]~\Theta\left(\frac{c}{M} -4\right),\label{M-func} 
 \end{equation}
 where $\Theta(x)$ is the Heaviside theta function and $\approx$ stands for logarithmic equivalent. Note that the special value $c = 4 M$ which 
 appears in Eq. (\ref{M-func}) is the upper (soft) edge associated to the Mar\v{c}enko-Pastur law \cite{MP67} which describes the density of eigenvalues of Wishart matrices as in Eq. (\ref{M-partition}) and without wall (i.e. $c \to \infty$). Therefore the theta function $\Theta(c/M - 4)$ appearing in Eq. (\ref{M-func}) is the zeroth order expression of the cumulative distribution of the largest eigenvalue of Wishart matrices in the large $M$ limit. Finally, injecting this expression (\ref{M-func}) of $\mathcal{M}_{M}\left(c,d\right)$ in Eq.~(\ref{h1}) and performing the $\tau$
 integration we get 
\begin{eqnarray}
\Delta_N &\simeq& W_N~\int_0^1dx~(2-x)^3 {x^{-(N^2+4)}}~\text{exp}\left[- \frac{2(N-2)(2-x)^2}{x^2}\right]~~\text{where} \label{h3} \\
 W_N &=& \frac{2^{N^2-N+6}}{(N-2)!}~(N-2)^{\frac{2N^2+N-2}{2}}~\text{exp}\left[-\frac{3}{4}(N-2)^2 \right] \approx N^{N^2}~~\text{for~large}~N.
 \label{W_N} 
\end{eqnarray}
It is straightforward to show that the remaining integral over $x$ in (\ref{W_N}) behaves for large $N$ as 
\begin{equation}
 \int_0^1dx~(2-x)^3 \frac{1}{x^{N^2+4}}~\text{exp}\left[- \frac{2(N-2)(2-x)^2}{x^2}\right] \simeq 
 \sqrt{\frac{\pi}{5}}\frac{N^{3/2}}{2^{N^2+3}}~\text{exp}\left[\frac{N^2}{2}\left(\log N-2\right)\right]\approx N^{\frac{N^2}{2}}
 ~~\text{for~large}~N.
\end{equation}
Hence 
\begin{equation}
\Delta_N = -h_N(0,0,...,0,-1) \approx N^{\frac{3N^2}{2}} = \text{exp}\left(\frac{3N^2}{2} \log N \right), 
\end{equation}
as given in Eq. (\ref{deltaN-lrgN}).


\begin{thebibliography}{}
\bibitem{Gumbel}E.~J. Gumbel, Statistics of Extremes,  Mineola, NY: Dover, ISBN 0-486-43604-7, (2004).
\bibitem{Dean}D.~S. Dean, S.~N. Majumdar, Extreme-value statistics of hierarchically correlated variables deviation from Gumbel statistics
and anomalous persistence, Phys. Rev. E {\bf 64}, 046121 (2001).
\bibitem{KrapivskySatya2003}S.~N. Majumdar, P.~L. Krapivsky, Extreme Value Statistics and Traveling Fronts: Various Applications, Physica A {\bf 318}, 161 (2003). 

\bibitem{Revuz99}D. Revuz, M. Yor, Continuous Martingales and Brownian Motion (Berlin: Springer), (1999).
\bibitem{Yor01}M. Yor, Exponential Functionals of Brownian Motion and Related Processes (Berlin: Springer), (2001).
\bibitem{Borodin02} A.~N.~Borodin, P. Salminen , Handbook of Brownian Motion: Facts and Formulae (Basel:Birkhauser), (2002).
\bibitem{Majumdar04}S. N. Majumdar, A. Comtet, {Exact Maximal Height Distribution of Fluctuating Interfaces}, Phys. Rev. Lett. {\bf 92}, 225501 (2004).

\bibitem{Majumdar05}S. N. Majumdar, A. Comtet, {Airy Distribution Function: From the Area Under a Brownian Excursion to the Maximal 
Height of Fluctuating Interfaces}, J. Stat. Phys. {\bf 119}, 777 (2005).

\bibitem{satyacurrsci}S. N. Majumdar, {Brownian Functionals in Physics and Computer Science},  Curr. Sci. {\bf{89}}, 2076 (2005).

\bibitem{Schehr10}G. Schehr, P. Le Doussal, {Extreme value statistics from the Real Space Renormalization Group: Brownian Motion, 
Bessel Processes and Continuous Time Random Walks}, J. Stat. Mech. P01009, (2010). 

\bibitem{Feller}W. Feller, An Introduction to Probability Theory and its Applications, (New York: Wiley), (1968).

\bibitem{Kearney04}M. J. Kearney, {{On a random area variable arising in discrete-time queues and compact directed percolation}}, 
J. Phys. A: Math. Gen. {\bf{37}}, 8421, (2004).
\bibitem{Kearney06}M. J. Kearney, {Exactly solvable cellular automaton traffic jam model}, Phys. Rev. E {\bf{74}}, 061115 (2006).

\bibitem{Comtet98}A. Comtet, C. Monthus, M. Yor, Exponential functionals of Brownian motion and disordered systems, J. Appl. Probab. {\bf{35}}, 255, (1998).
\bibitem{majum-Qfin}S. N. Majumdar, J.-P. Bouchaud, {{Optimal time to sell a stock in 
the Black-Scholes model: comment on "Thou shalt buy and hold"}}, by A. Shiryaev, Z. Xu and X.Y. Zhou,  
Quant. Fin., {\bf{8}}, 753 (2008).


\bibitem{Bramson91} M. Bramson, D. Griffeath, Capture problems for coupled random walks, in Random
Walks, Brownian Motion and Interacting Particle Systems, R. Durrett and H. Kesten,
eds. 153-188. Birkha\"user, Boston (1991).
\bibitem{Wenbo01}W. V. Li, Q.-M. Shao, Capture time of Brownian pursuits, Probab. Theory Rel. {\bf 121}, 30 (2001). 

\bibitem{Blumen84}A. Blumen, G. Zumofen, J. Klafter, {Target annihilation by random walkers}, Phys. Rev. B {\bf 30}, 5379 (1984).

\bibitem{Kang85} K. Kang, S. Redner, Fluctuation-dominated kinetics in diffusion-controlled reactions, Phys. Rev. A {\bf 32}, 435 (1985).
\bibitem{Ben-Naim93}E. Ben-Naim, S. Redner, F. Leyvraz, Decay kinetics of ballistic annihilation, Phys. Rev. Lett. {\bf 70}, 1890 (1993). 
\bibitem{Krapivsky95}P.~L. Krapivsky, S. Redner, F. Leyvraz, Ballistic annihilation kinetics: The case of discrete velocity distributions, 
Phys. Rev. E {\bf 51}, 3977 (1995).

\bibitem{Bray94} A. J. Bray,  Theory of phase-ordering kinetics, Adv. Phys. {\bf 43}, 357 (1994).

\bibitem{Shiryaev} A. N. Shiryaev, Optimal Stopping Rules. Springer, ISBN 3-540-74010-4, (2007). 



\bibitem{Rednerbook} S. Redner, A guide to first-passage processes, Cambridge University Press, Cambridge, (2001).
\bibitem{Bray13}A. J. Bray, S. N. Majumdar, G. Schehr, {{Persistence and First-Passage Properties in Non-equilibrium Systems}}, 
Adv. Phys {\bf{62}}, 225 (2013).


\bibitem{Kearney05}M. J. Kearney, S. N. Majumdar, {On the area under a continuous time Brownian motion till its first-passage time}, 
J. Phys. A: Math. Gen. {\bf{38}}, 4097 (2005).

\bibitem{Satyacurrsci99}S. N. Majumdar, Persistence in nonequilibrium systems, Curr. Sci. {\bf 77}, 370 (1999).
\bibitem{Satyarosso10}S. N. Majumdar, A. Rosso, A. Zoia, Hitting Probability for Anomalous Diffusion Processes, Phys. Rev. Lett. {\bf 104}, 020602 (2010).



\bibitem{Kearney07}M. J. Kearney, S. N. Majumdar, R. J. Martin, {The first-passage area for drifted Brownian motion and the moments 
of the Airy distribution}, J. Phys. A: Math. Theor. {\bf{40}}, F863 (2007).
\bibitem{Randon-Furling07}J. Randon-Furling, S. N. Majumdar, {Distribution of the time at which the deviation of a Brownian motion is maximum 
before its first-passage time}, J. Stat. Mech. P10008, (2007).
\bibitem{Abundo13}M. Abundo, {On the First-Passage Area of a One-Dimensional Jump-Diffusion Process}, Methodol. Comput. Appl. {\bf{15}}, 85 (2013).



\bibitem{SatyaBray2010} S. N. Majumdar, A. J. Bray, {Maximum distance between the Leader and the Laggard for three Brownian walkers}, 
J. Stat. Mech. {{P08023}}, (2010).

\bibitem{KrapivskySatya2010}P. Krapivsky, S. N. Majumdar, A. Rosso, {Maximum of N independent Brownian walkers 
till the first exit from the half-space}, J. Phys. A: Math. Theor. {\bf{43}}, 315001 (2010).

\bibitem{deGennes68}P. G. de Gennes, Soluble Model for Fibrous Structures with Steric Constraints, J. Chem. Phys. {\bf 48}, 2257 (1968).

\bibitem{Fisher84}M. E. Fisher, Walks, walls, wetting, and melting, J. Stat. Phys. {\bf 34}, 667 (1984).






\bibitem{Schehr08}G. Schehr, S. N. Majumdar, A. Comtet, J. Randon-Furling, 
Exact Distribution of the Maximal Height of $p$ Vicious Walkers,  Phys. Rev. Lett. {\bf{101}}, 150601 (2008).
\bibitem{Kobayashi}N. Kobayashi, M. Izumi, M. Katori, {Maximum distributions of bridges of noncolliding Brownian paths},
Phys. Rev. E {\bf 78}, 051102 (2008). 
\bibitem{Nadal09} C. Nadal, S. N. Majumdar, Nonintersecting Brownian interfaces and Wishart random matrices, Phys. Rev. E {\bf 79}, 061117 (2009).
\bibitem{Forrester08}P. J. Forrester, S. N. Majumdar, G. Schehr, 
Non-intersecting Brownian walkers and Yang-Mills theory on the sphere, Nucl. Phys. B {\bf{844}}, 500 (2011). 
\bibitem{Izumi11} M. Izumi, M. Katori, Extreme value distributions of noncolliding diffusion processes (math.PR/1006.5779), 
                RIMS Kokyuroku Bessatsu B27, 45 (2011).
\bibitem{Rambeau11}J. Rambeau, G. Schehr, {Distribution of the time at which N vicious walkers reach their maximal height}, 
Phys. Rev. E {\bf{83}}, 061146 (2011).
\bibitem{Issac}I. Perez-Castillo, T. Dupic, Reunion probabilities of $N$ one-dimensional random walkers with mixed boundary conditions, 
arXiv:1311.0654, (2013).



\bibitem{Anupam13}A. Kundu, S. N. Majumdar, G. Schehr, Exact Distributions of the Number of Distinct and Common Sites Visited 
by N Independent Random Walkers, Phys. Rev. Lett. {\bf{110}}, 220602 (2013).
\bibitem{Gabel12}A. Gabel, S. N. Majumdar, N. K. Panduranga, S. Redner, Can a Lamb Reach a Haven Before Being Eaten by Diffusing Lions?, 
J. Stat. Mech. P05011, (2012).





%
%
%





\bibitem{Schaums} M. R. Spiegel, S. Lipschutz, J. J. Schiller, D. Spellman, {Schaum's outlines: Complex Variables}, Second Edition, 
McGraw-Hill, (2009).



\bibitem{karlin}S. Karlin, J. McGregor, Coincidence probabilities, Pacific J. Math. {\bf 9}, 1141 (1959).
\bibitem{Krattenthaler} C. Krattenthaler, A. J. Guttmann, X. G. Viennot, Vicious walkers, friendly walkers and Young tableaux: 
II. With a wall, J. Phys. A: Math. Gen. {\bf 33}, 8835 (2000).
\bibitem{Bray04} A. J. Bray, K. Winkler, {Vicious walkers in a potential}, J. Phys. A: Math. Gen. {\bf{37}}, 5493 (2004).


\bibitem{Dyson}F. J. Dyson, Statistical Theory of the Energy Levels of Complex Systems. I, J. Math. Phys. {\bf 3}, 140 (1962).

\bibitem{Dean08} D. S. Dean, S. N. Majumdar, Large Deviations of Extreme Eigenvalues of Random Matrices, Phys. Rev. Lett. {\bf 97}, 160201 (2006).
                
\bibitem{Dean08II} D. S. Dean, S. N. Majumdar, Extreme value statistics of eigenvalues of Gaussian random matrices, Phys. Rev. E {\bf 77}, 041108 (2008).                
                
\bibitem{Bohigas}P. Vivo, S. N. Majumdar, O. Bohigas, Large deviations of the maximum eigenvalue in Wishart random matrices, 
J. Phys. A: Math. Theor. {\bf 40}, 4317 (2007).
      
\bibitem{Majumdar14}S. N. Majumdar, G. Schehr, Top eigenvalue of a random matrix: large deviations and third order phase transition, J. Stat. Mech. P01012 (2014).       

\bibitem{Rambeau10}J. Rambeau, G. Schehr, Extremal statistics of curved growing interfaces in 1+1 dimensions, Europhys. Lett. {\bf 91}, 60006 (2010).



\bibitem{Mehta}
M. L. Mehta, \emph{Random Matrices}, 2nd Edition, Academic Press (1991).


\bibitem{Forrester}
P.~J. Forrester, \emph{Log-gases and random matrices}, Princeton University
  Press, Princeton, NJ, (2010).

\bibitem{MS14}
S. N. Majumdar, G. Schehr, Top eigenvalue of a random matrix: large deviations and third order phase transition, J. Stat. Mech. P01012 (2014). 


\bibitem{MP67}
V. A. Mar\v{c}enko, L. A. Pastur, Distribution of eigenvalues in certain sets of random matrices, Math. USSR-Sb. {\bf 1}, 457 (1967). 


\end{thebibliography}
\end{document}